\documentclass[sigconf]{acmart}

\usepackage{multirow}
\usepackage{xcolor}
\usepackage{tabularray}
\usepackage{hyperref}
\usepackage{subcaption}

\usepackage{framed}
\usepackage[most]{tcolorbox}
\newtcolorbox{bluebox}[1][]{
  colback=blue!5!white,
  colframe=blue!75!black,
  boxsep=2pt,      
  top=2pt,         
  bottom=2pt,      
  left=5pt,        
  right=5pt,       
  width=\linewidth 
}

\usepackage{blindtext}
\usepackage{balance}
\balance

\definecolor{main}{HTML}{5989cf}    
\definecolor{sub}{HTML}{cde4ff}     

\definecolor{firebrick}{HTML}{B22222}
\definecolor{yellowgreen}{HTML}{9ACD32}
\definecolor{royalblue}{HTML}{4169E1}

\begin{document}
\title[Increasing, not Diminishing]{Increasing, not Diminishing: Investigating the Returns of Highly Maintainable Code}

\author{Markus Borg}
\orcid{XXX}
\affiliation{%
  \institution{CodeScene and Lund University}
  \city{Malm\"o}
  \country{Sweden}
}
\email{markus.borg@codescene.com}

\author{Ilyana Pruvost}
\orcid{XXX}
\affiliation{%
  \institution{University of Clermont Auvergne}
  \city{Clermont-Ferrand}
  \country{France}
}
\email{ilyana.pruvost@etu.uca.fr}

\author{Enys Mones}
\orcid{XXX}
\affiliation{%
  \institution{CodeScene}
  \city{Malm\"o}
  \country{Sweden}
}
\email{enys.mones@codescene.com}

\author{Adam Tornhill}
\orcid{XXX}
\affiliation{%
  \institution{CodeScene}
  \city{Malm\"o}
  \country{Sweden}
}
\email{adam.tornhill@codescene.com}

\renewcommand{\shortauthors}{Borg \textit{et al.}}

\begin{abstract}
Understanding and effectively managing Technical Debt (TD) remains a vital challenge in software engineering. While many studies on code-level TD have been published, few illustrate the business impact of low-quality source code. In this study, we combine two publicly available datasets to study the association between code quality on the one hand, and defect count and implementation time on the other hand. We introduce a value-creation model, derived from regression analyses, to explore relative changes from a baseline. Our results show that the associations vary across different intervals of code quality. Furthermore, the value model suggests strong non-linearities at the extremes of the code quality spectrum. Most importantly, the model suggests amplified returns on investment in the upper end. We discuss the findings within the context of the ``broken windows'' theory and recommend organizations to diligently prevent the introduction of code smells in files with high churn. Finally, we argue that the value-creation model can be used to initiate discussions regarding the return on investment in refactoring efforts.
\end{abstract}

\copyrightyear{2024}
\acmYear{2024}
\setcopyright{acmlicensed}\acmConference[TechDebt 2024]{7th International Conference on Technical Debt}{April 14--15, 2024}{Lisbon, Portugal}
\acmPrice{15.00}
\acmDOI{10.1145/3593434.3593480}
\acmISBN{979-8-4007-0044-6/23/06}


\maketitle

\section{Introduction}
Efficient and effective Technical Debt (TD) management is a critical challenge for large software projects. The consequences of neglecting TD are highlighted by the recent case of Southwest Airlines in 2022~\cite{tufekci_shameful_2022}. The company's debt-ridden crew scheduling system bulged under unexpected winter weather disruptions, leading to extensive flight cancellations and significant financial losses --- approximately \$410 million. This example underlines that TD remains a vital concern in the software industry.

A large body of academic research emphasizes many potential risks with TD. Not only are secondary studies available, but there are even tertiary studies~\cite{rios_tertiary_2018,junior_consolidating_2022}. Junior and Travassos report that academic research targeted TD of various nature, i.e., source code, design, architecture, build systems, documentation, requirements, tests, services, and versioning~\cite{junior_consolidating_2022}. They also list nine concrete negative impacts identified in the academic literature. Example impacts include hampered maintainability and evolution, decreased internal quality, additional defects, reduced code comprehension, decreased productivity, higher organizational costs, and lower developer morale.

Most research focuses on code-level TD. While this type of debt might be easier to identify using analysis tools, the broader business implications of the tool output are less explored. We previously published a notable exception~\cite{tornhill_code_2022}, reporting statistically significant differences between low- and high-quality code. More specifically, we found that low-quality code (compared to high-quality code) on average 1) exhibits 15 times more defects, 2) requires more than twice as long time for issue resolution, and 3) faces greater uncertainty in issue resolution times. However, our previous work only presented a course-grained analysis of three categories of code quality, measured using CodeScene's Code Health metric.

Now we conduct a regression analysis of the study's publicly available data~\cite{tornhill_code_2022}. Our analysis provides deeper insights into the relationship between code quality, defect counts, and issue resolution time. Moreover, we extend the dataset with another publicly available dataset with a similar structure~\cite{borg_u_2023}. In total, our analysis is based on data from 79~proprietary software projects from various companies. For a study restricted to closed-source software, the size and diversity of this dataset can be considered very large.  

Defect counts and development velocity are key factors in a software organization's value creation. Although software productivity is a multi-faceted and complex concept, as discussed by Forsgren \textit{et al.}~\cite{forsgren_space_2021}, there is no question that fast-moving developers producing defect-free code is valuable --- this is close to the ``better software faster''~\cite{farley_modern_2021} core of software engineering. This paper proposes a pragmatic value-creation model that merges these two critical factors. Our goal is to spark discussions on the business value associated with varying levels of code quality. 

Two questions guide our research:
\begin{itemize}
\item[RQ1] How is Code Health associated with different levels of a) defect count and b) Time-in-Development?
\item[RQ2] How do different levels of Code Health translate into relative differences in value creation?
\end{itemize}

RQ1 expands on the findings of previous work, as we go beyond the statistically significant different averages to explore more fine-grained relationships between the variables. For RQ2, our proposed value-creation model synthesizes the insights from RQ1's regression models. Furthermore, we discuss our findings from the lens of the ``broken windows'' theory~\cite{hunt_pragmatic_1999}, recently confirmed by Levén \textit{et al.} in the TD context~\cite{leven_testing_2023}. The theory captures the psychological mechanisms that drive developers' approaches to code maintainability, especially at the extremities of code quality. For pristine quality, the theory explains how developers vigilantly avoid introducing the first code smell. For abysmal quality, the theory suggests that developers might care less about code that is already plagued by numerous smells.

Our findings reveal that the greatest variations in defect counts are observed at the lower end of the Code Health spectrum. Time-in-Development involves more uncertainty, yet the overall trend displays faster development for higher levels of Code Health. However, the highest Time-in-Development is not observed for the lowest levels of Code Health. This could possibly be explained by the broken windows theory, as developers might be more inclined to add subpar fixes in already terribly bad code. More prominently, our proposed value-creation model suggests a heightened sensitivity to Code Health changes in the uppermost Code Health range. Any introduction of code smells in a project's central files --- files with high churn, i.e., high TD interest rates --- could potentially have a direct business impact on a project.

The paper is organized as follows. Sec.~\ref{sec:bg} introduces code quality, the measures we use, and the broken windows theory. Sec.~\ref{sec:rw} shares related work on relationships between code smells, defect count, and maintenance effort. Sec.~\ref{sec:method} describes the dataset creation and details our analysis approach. In Sec.~\ref{sec:res} we present our results and discuss their implications. Sec.~\ref{sec:limit} discusses the limitations of our value-creation model and Sec.~\ref{sec:threats} elaborates on the threats to validity. Finally, we conclude the paper and outline future work in Sec.~\ref{sec:conc}.

\section{Background} \label{sec:bg}
This section describes the two primary measures used in this study and introduces the broken windows theory.

\subsection{Code Health and Time-in-Development}
Several scholars have put forward metrics to assess source code quality, resulting in an extensive array of options in this field~\cite{riaz_systematic_2009,baggen_standardized_2012}. Nunez-Varela \textit{et al.} conducted a comprehensive review in 2017, revealing close to 300 different metrics~\cite{nunez-varela_source_2017}. The predominant focus in academic circles has been on object-oriented programming, with 1) the Chidamber and Kemerer metrics, 2) Lines of Code (LoC), 3) McCabe's cyclomatic complexity, and 4) the tally of methods and attributes emerging as the most prevalent metrics. The ISO/IEC~5055 standard, released in March 2021, introduces a novel approach to evaluating system attributes such as maintainability, reliability, security, and performance directly at the source code level, eliminating the need for system execution. This standard achieves its purpose by counting occurrences from a defined list of detrimental (low-level) weaknesses~\cite{international_organization_for_standardization_information_2021}.

\textbf{Code Health} (CH) is quantified through the CodeScene tool, yielding a numeric and absolute value that ranges between 10 (top-notch code quality) and 1 (extremely poor code quality). This metric aligns with the philosophy that the optimal strategy for gauging source code complexity is to pinpoint and quantify specific complexity attributes~\cite{fenton_software_1994}. To realize this, CH integrates elements from the maintainability category of ISO/IEC~5055 and supplements them with design-level code smells, which include but are not limited to God Class, God Methods, and Duplicated Code~\cite{lacerda_code_2020}. The resulting CH leads to the categorization of files into one of three sub-intervals: healthy ($CH \geq 9$), warning ($4 \leq CH < 9$), and alert ($ CH < 4$). In this study, we use CH as the independent variable.

In line with previous work~\cite{tornhill_code_2022,borg_u_2023}, our analysis combines CH with the duration of development cycles. In terms used by Jira, cycle time refers to the duration from when an action is initiated to its completion, marking the period when the issue is tagged as ``In progress.'' Contrastingly, lead time encompasses the total time from the submission of an issue to its closing. In our work, we choose cycle time and refer to it as \textbf{Time-in-Development} (Time-in-Dev)~\cite{tornhill_code_2022}. This time measure uses a combination of Jira's transition timestamps and git commit metadata. By focusing on Time-in-Dev, we exclude time spent in post-development activities like testing, release management, and deployment. 
We use Time-in-Dev as a dependent variable, complemented by defect counts --- defined as the number of Jira issues of type `bug'.

\subsection{The Broken Windows Theory}
In 1982, Wilson and Kelling proposed the ``broken windows'' theory in an opinion piece in The Atlantic~\cite{wilson_broken_1982}. They hypothesized that disorder in urban neighborhoods encourages more serious crime. The ideas gradually evolved into a criminological theory that has shaped policing strategies in various locations, e.g., a well-known zero-tolerance toward graffiti and public drinking in Manhattan. The theory suggests that if disorders such as broken windows go unpunished, others might be encouraged to violate laws and norms. 


The broken windows theory has influenced several other research fields. For example, it has influenced research on substance abuse in public health care. In business management, it has been used to recommend small problems before they escalate. A third example is how the theory has been used to influence educational policies to quickly address minor behavioral issues. 

The theory transposed into software engineering through the popular book ``The Pragmatic Programmer’’ by Hunt and Thomas~\cite{hunt_pragmatic_1999}. Their main point was that small problems in a codebase, e.g., sloppy coding practices and quick fixes, could trigger further quality deterioration. The authors strongly recommend that development teams address issues promptly as they are identified and cultivate a culture of quality. For the really poor code, they warn that developers might ignore best practices and just follow suit. However, Hunt and Thomas based the discussion on experience without any empirical evidence.

Lev\'en \textit{et al}. recently confirmed the theory through a controlled experiment~\cite{leven_testing_2023}. Twenty-nine developers extended software characterized by either high or low TD levels. Based on SonarQube measurements, they found that the presence of existing TD heightened the propensity for developers to introduce additional debt during system extensions. Furthermore, the authors did a causal analysis (based on a simple four-node directed acyclic graph) from which they inferred the causality of pre-existing TD on accumulating even more TD. 

\section{Related Work} \label{sec:rw}
Many previous studies correlate code smells with various software engineering aspects.
Santos \textit{et al.} conducted a systematic literature review to analyze what effects of code smells have been reported in empirical studies~\cite{santos_systematic_2018}. Part of their contribution was to summarize previous work on correlational studies. They found that the two most studied correlations were 1) changes and defects and 2) maintenance effort. The next subsections present selected studies concerning these aspects.

\subsection{Code Smells and Defect Count}
Santos \textit{et al.} identified 26 primary studies focusing on correlations with changes (i.e., maintenance or refactoring activities) and defects~\cite{santos_systematic_2018}. While change and defect count are related concepts, we focus on summarizing research on defects.

Li and Shatnawi analyzed post-release defects for three versions of Eclipse (2.0, 2.1, and 3.0)~\cite{li_empirical_2007}. They found that three code smells (\textit{Shotgun Surgery}, \textit{God Class}, and \textit{God Methods}) were associated with higher defect proneness. Zhang \textit{et al.} studied the impact of bad design indicators on file-level defect density on 18 release versions of the Apache Commons series~\cite{zhang_empirical_2017}. They found that files with these indicators are significantly more likely to be faulty. More recently, we studied source code from 39 proprietary projects and found that files with low CH are 15 times more defect prone~\cite{tornhill_software_2018}. 

The body of literature on the correlation between code smells, and other code measures, has inspired numerous researchers to work on defect prediction~\cite{hall_systematic_2012,radjenovic_software_2013,hosseini_systematic_2019}. Our work is partly inspired by such work, although we do not aim to predict the risks of identifying defects in individual files.  We examine overarching trends, exploring how shifts in code quality at the file level can influence the average number of defects.

\subsection{Code Smells and Maintenance Effort}
Santos \textit{et al.} identified six primary studies linking code smells with increased maintenance effort, highlighting the challenges they pose to maintenance tasks~\cite{santos_systematic_2018}.

Olbrich \textit{et al.} investigated how the two code smells \textit{God Class} and \textit{Shotgun Surgery} were associated with development efforts in two large-scale open-source systems~\cite{olbrich_evolution_2009}. They concluded that classes ``infected'' by the code smells need more maintenance than non-infected classes, i.e., \textit{God Class} leads to bigger changes and \textit{Shotgun Surgery} triggers more frequent maintenance changes. Zazworka \textit{et al.} also studied the \textit{God Class} and drew similar conclusions~\cite{zazworka_investigating_2011}. Based on a small study of a proprietary development project, they report that \textit{God Classes} require more maintenance effort by requiring more changes, more changes related to fixing bugs, and more effort per change.

Yamashita and Moonen studied maintainability aspects of four Java information systems with feature parity~\cite{yamashita_code_2012}. Their empirical study identified a correlation between maintainability factors and code smells. They also report that code smell detection can cover a more heterogeneous spectrum of maintainability factors than software metrics and expert judgment individually. In line with these results, the datasets analyzed in our current study are based on code smells rather than low-level code metrics.

Sj{\o}berg \textit{et al.} studied the relationship between 12 code smells and maintenance effort~\cite{sjoberg_quantifying_2013}. They hired six developers to perform maintenance tasks on four functionally equivalent Java systems (the same systems that Yamashita and Moonen studied). Based on a regression analysis, the authors concluded that the effects of the smells were limited. Instead, they recommend a focus on limiting code size. However, Soh \textit{et al.} analyzed the same data using a more fine-grained analysis of maintenance sub-tasks~\cite{soh_code_2016}, i.e., reading, editing, searching, and navigating source code. They found that code smells indeed are more detrimental than file size when editing and navigating code. The approach of re-analyzing previous data motivated our work in this paper. 

Our previous mining study also explored the relationship between code smells and effort in terms of development time~\cite{tornhill_code_2022}. We reported that resolving issues in low-quality code is on average 124\% slower than in high-quality code. In this study, we perform a more thorough analysis of an extended version of this dataset.

\section{Method} \label{sec:method}
We rely on previous repository mining studies that extracted data from source code repositories and issue management systems~\cite{chatterjee_empirical_2022}. In this paper, we combine data and perform novel quantitative analyses. Our study is observational in nature, meaning that we do not introduce any interventions; instead, our primary focus is on identifying meaningful patterns~\cite{ayala_use_2022}.

\subsection{Data Collection and Preprocessing} \label{sec:data}
We combine publicly available data from two previous (disjoint) studies of CodeScene customers~\cite{tornhill_code_2022,borg_u_2023} --- referred to as the \textit{CodeRed} dataset and the \textit{UOwns} dataset, respectively. The former dataset is organized per source code file whereas the latter is based on touches, i.e., individual modifications to single files~\cite{foucault_code_2014}. Based on information from the original publications, we characterize the included software projects as follows:

\begin{itemize}
    \item[CodeRed] 39 proprietary development projects with an average codebase containing 171 kLoC. Several domains are represented, e.g., retail, construction, infrastructure, brokers, and data analytics. The files' CH distribution is 89.0\% healthy, 9.9\% warning, and 1.1\% alert. Roughly 130,000 Jira issues are represented in the dataset, in which C++ is the most common language.
    \item[UOwns] 40 proprietary development projects containing 139,869 files in total (including non-source code) contributed by 1,414 developers. The source code files' CH distribution is 84.5\% healthy, 14.5\% warning, and 1.0\% alert. About 30,000 Jira issues were collected. C\# is the most common language in the dataset.
\end{itemize}

The merged dataset contains rows of changes to unique files. 
All changes occurred during the time window used for CodeScene analyses (by default 1~year). This ensures that the collected data relates to recent, and thus relevant, development activities. We followed similar preprocessing steps as described in the original papers~\cite{tornhill_code_2022,borg_u_2023}. For CodeRed, this includes removing Time-in-Dev outliers that are more than 1.5 $\times$ interquartile range above the third quartile or below the first quartile. For UOwns, we removed outliers representing touches for which Time-in-Dev or change size is not covered by three standard deviations. Finally, we restructured UOwns to the file-based format of CodeRed and combined all data. The details can be found in the accompanying replication package~\cite{notebook}.

Table~\ref{tab:desc_stats} shows an overview of the dataset analyzed in this study. In total, our combined dataset used for this study contains 46,211 source code files. The table presents the average, lower quartile (25\%), median (50\%), upper quartile (75\%), and standard deviation (StD) for the following metrics collected during the time window: 1) CH at the time of the most recent commit, 2) the number of Jira issues of type ``bug'' connected to the file, and 3) the total amount of Time-in-Dev for Jira issues connected to the file. We note that the dataset is heavily skewed toward files with high Code Health scores and few defects. Moreover, the standard deviation for both defect counts and Time-in-Dev is very large.

\begin{table}[]
\caption{Descriptive statistics of the files in the dataset.}
\begin{tabular}{l|c|c|c|c|c|}
\cline{2-6}
                                 & Average & 25\%  & 50\%  & 75\%  & StD    \\ \hline
\multicolumn{1}{|l|}{CH}         & 9.60    & 9.75  & 10    & 10    & 1.07   \\ \hline
\multicolumn{1}{|l|}{\#Defects}  & 0.50    & 0     & 0     & 0     & 2.09   \\ \hline
\multicolumn{1}{|l|}{Time (min)} & 7,501   & 1,380 & 3,360 & 7,680 & 19,809 \\ \hline
\end{tabular}
\label{tab:desc_stats}
\end{table}

\begin{figure}
    \centering
    \includegraphics[width=0.45\textwidth]{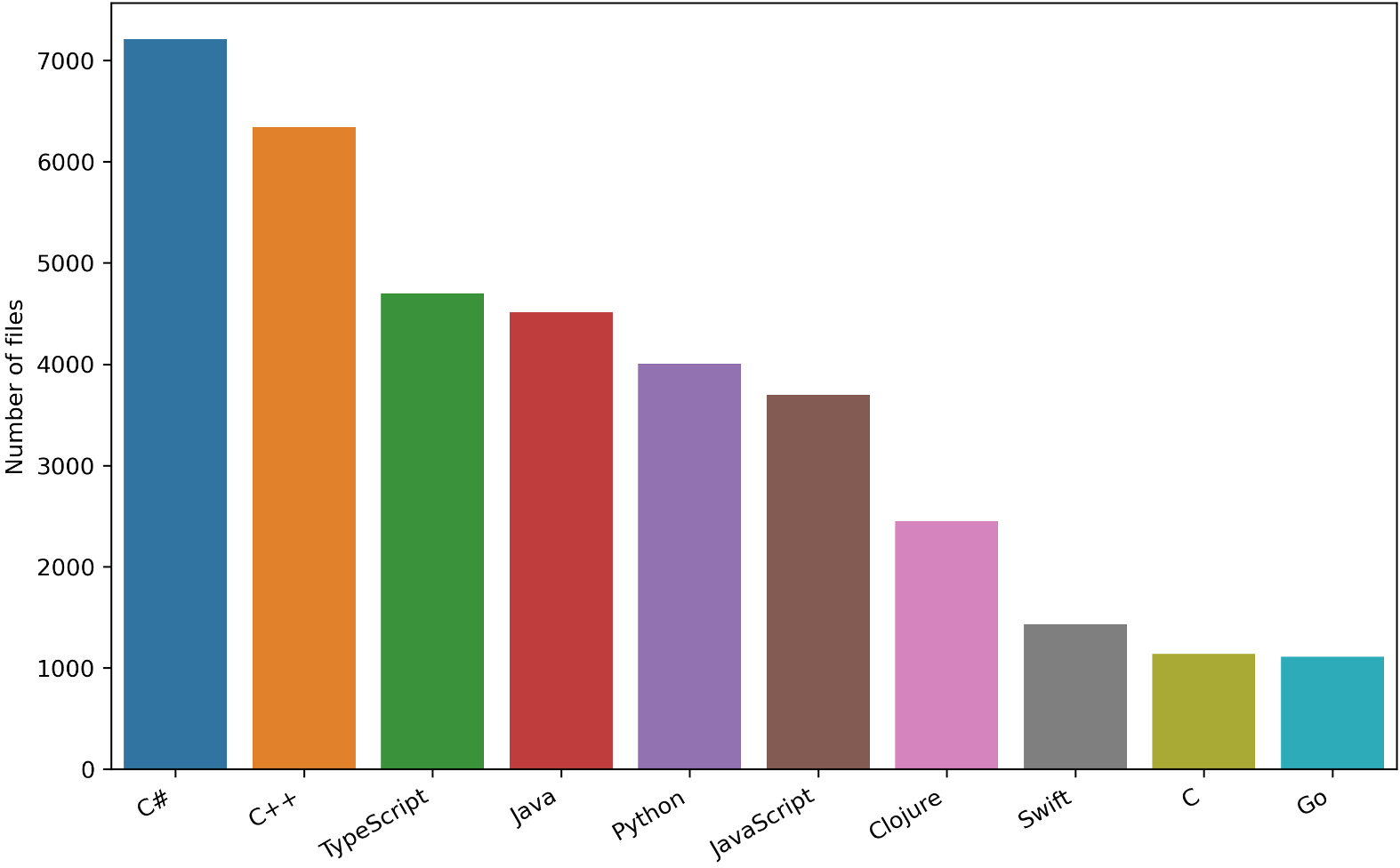}
    \caption{Top-10 programming languages in the dataset.}
    \label{fig:programminglanguages}
\end{figure}

Fig.~\ref{fig:programminglanguages} shows the distribution of the most common languages. There are 25 programming languages represented in the dataset. Predominantly, it showcases the most popular languages in the industry occupying the top ranks. When JavaScript and TypeScript are combined, they emerge as the most represented languages. C\# and C/C++ share roughly an equal level of representation. Additionally, the dataset includes a substantial number of files in Java and Python, each exceeding 4,000 files.

\subsection{Data Analysis} \label{sec:analysis}
This subsection describes the regression modeling and our proposed value-creation model.

\subsubsection{Regression Models} \label{sec:regression}
In our study, we seek to model relationships between CH as the independent variable, and two dependent variables: 1) average number of defects (Defects) and 2) average Time-in-Dev. Fitting regression models to complex software engineering data is challenging and the combined dataset unfortunately does not give us insights into some of the key confounders that are likely to influence the dependent variables. However, the substantial size of our dataset allows for the identification of meaningful patterns, even with regression models of limited predictive power.

Our regression analysis examined polynomial degrees ranging from one and four. Higher degrees are likely to result in overfitting and would be difficult to explain using our domain understanding and existing theoretical models. We anticipated that the relationships between the variables would need more expressivity beyond that of a simple linear model. As suggested by the broken windows theory, there might be non-linear patterns at the extremes of the quality spectrum.

Table~\ref{tab:reg_eval} presents results from the polynomial degree selection for Defects and Time-in-Dev. $R^{2}$ indicates the proportion of the variance in the dependent variable that is predictable from the independent variables. MAE is the average of the absolute differences between predictions and actual observations. Finally, MSE is the average of the squares of the residuals. We find that none of the models capture a large fraction of the variability. The third-degree model (presented in bold font) is slightly better than the alternatives for Defects. Based on the non-linearity argumentation above, we rely on it for the remainder of this work. The differences for Time-in-Dev are even smaller. For consistency, and following the same non-linearity rationale, we proceed with the third-degree model.

\begin{table}[]
\caption{Regression model performance metrics.}
\begin{tabular}{l|lll|lll|}
\cline{2-7}
                                     & \multicolumn{3}{c|}{Defects}                                                                         & \multicolumn{3}{c|}{Time-in-Dev}                                                            \\ \hline
\multicolumn{1}{|l|}{Degree}         & \multicolumn{1}{c|}{$R^{2}$}             & \multicolumn{1}{c|}{MAE}            & \multicolumn{1}{c|}{MSE} & \multicolumn{1}{c|}{$R^{2}$}             & \multicolumn{1}{c|}{MAE}           & \multicolumn{1}{c|}{MSE} \\ \hline
\multicolumn{1}{|l|}{First}          & \multicolumn{1}{l|}{0.032}          & \multicolumn{1}{l|}{0.756}          & 4.227                    & \multicolumn{1}{l|}{0.005}          & \multicolumn{1}{l|}{7203}          & $3.9 \times 10^{8}$                    \\ \hline
\multicolumn{1}{|l|}{Second}         & \multicolumn{1}{l|}{0.032}          & \multicolumn{1}{l|}{0.757}          & 4.226                    & \multicolumn{1}{l|}{0.006}          & \multicolumn{1}{l|}{7209}          & $3.9 \times 10^{8}$                    \\ \hline
\multicolumn{1}{|l|}{\textbf{Third}} & \multicolumn{1}{l|}{\textbf{0.038}} & \multicolumn{1}{l|}{\textbf{0.748}} & \textbf{4.199}           & \multicolumn{1}{l|}{\textbf{0.006}} & \multicolumn{1}{l|}{\textbf{7209}} & $\mathbf{3.9 \times 10^{8}}$     \\ \hline
\multicolumn{1}{|l|}{Fourth}         & \multicolumn{1}{l|}{0.038}          & \multicolumn{1}{l|}{0.749}          & 4.199                    & \multicolumn{1}{l|}{0.006}          & \multicolumn{1}{l|}{7209}          & $3.9 \times 10^{8}$                   \\ \hline
\end{tabular}
\label{tab:reg_eval}
\end{table}

To address the poor predictability of our regression models, we evaluate the uncertainty of their regression coefficients using bootstrapping~\cite{efron_bootstrap_1979}. Bootstrapping is a statistical technique used to estimate the reliability of data analyses. The technique involves repeatedly resampling data --- with replacement --- to assess how various parameters might vary without making strict statistical assumptions. In this study, we run bootstrapping 10,000 times and each bootstrap sample has the same size as the total dataset. This approach preserves the original distribution of the data. For each bootstrap sample, we calculate the corresponding regression coefficients. The resulting plots bring visual insights into the uncertainty as we can study how the regression lines vary when different samples are taken from the dataset.

Kläs \textit{et al.} have previously proposed using bootstrapping for defect prediction using hybrid models~\cite{klas_handling_2011}. Our work shows another application in the software engineering context.

\subsubsection{Value-creation Model} \label{sec:value}
We propose a simple value-creation model to discuss the relationship between CH and the value created by a software organization. The model is inspired by two metrics of the SPACE framework proposed by Forsgren \textit{et al.}~\cite{forsgren_space_2021}, namely \textit{coding time} from the activity dimension and \textit{lack of interruption} from the efficiency and flow dimension. In a nutshell, the model multiples two factors:
\begin{itemize}
    \item \textbf{Capacity} ($c$). An organization's capability to produce valuable output, i.e., the maximum output the organization can achieve under ideal conditions. This can include factors like the number of developers, tools available, and infrastructure. From the SPACE perspective, a higher capacity represents more coding time.
    \item \textbf{Efficiency} ($e$). How effectively the organization can convert its capacity into actual value. The factor takes into account wastages, bottlenecks, defects, etc. Efficiency reflects ``lack of interruption'' in SPACE.
\end{itemize}

We model value creation as a product of capacity and efficiency. Furthermore, we assume that efficiency corresponds to 1 minus the \textbf{fraction of unplanned work} ($u$), i.e., $1 - u$. We combine the above discussion into a pragmatic model of a software organization's value creation $v$ as shown in Equation~\ref{eq:value}.

\begin{equation}
\text{v} = \text{c} \times (1-u)
\label{eq:value}
\end{equation}

Moreover, we posit two proportionalities in the context of our value-creation model. First, \textit{capacity is inversely proportional to Time-in-Dev} ($t$) (cf. Equation~\ref{eq:prop-time}). The rationale is that the longer developers work on implementing a resolution, the more resources are tied up addressing it. These resources could instead have been allocated to new feature development that contributes directly to the development capacity.

\begin{equation}
\text{c} \propto \frac{1}{t}
\label{eq:prop-time}
\end{equation}

Second, the \textit{fraction of unplanned work is proportional to the number of defects} ($d$) (cf. Equation~\ref{eq:prop-defects}). Our motivation is that defects divert developer time from planned work on new features to other tasks. Every defect necessitates unplanned work to diagnose, repair, test, and redeploy the software. Furthermore, defect management might lead to communication overheads, both within the team and with customers.

\begin{equation}
\text{u} \propto \text{d}
\label{eq:prop-defects}
\end{equation}

Finally, we use the value-creation model and the proportionalities to illustrate what different levels of CH represent in terms of relative changes. Output from the regression models described in Sec.~\ref{sec:regression} is used as input to the value-creation model. Our discussions are based on a given \textit{starting point} representing an initial CH ($CH_{0}$) and a fraction of unplanned work ($u_{0}$) which yields an average Time-in-Dev for implementing the resolution for a Jira issue and an average number of defects. Using the same regression models, we calculate the relative differences representing other levels of CH. 
As a last step, we plot the value creation representing the starting point and what other levels of CH would yield given Equation~\ref{eq:value}. We elaborate on the limitations of this approach in Sec.~\ref{sec:limit}.

Our plots also illustrate the uncertainty of the value function. To this end, we use the covariance matrices from the regression models to estimate the joint probability distribution of the regression coefficients. We then sample the regressed polynomial from this distribution 10,000 times and plot the results.

\section{Results and Discussion} \label{sec:res}
This section reports our results and discusses their implications.

\subsection{Regression models (RQ1)} \label{sec:res-rq1}
\begin{figure}
    \centering
    \includegraphics[width=0.45\textwidth]{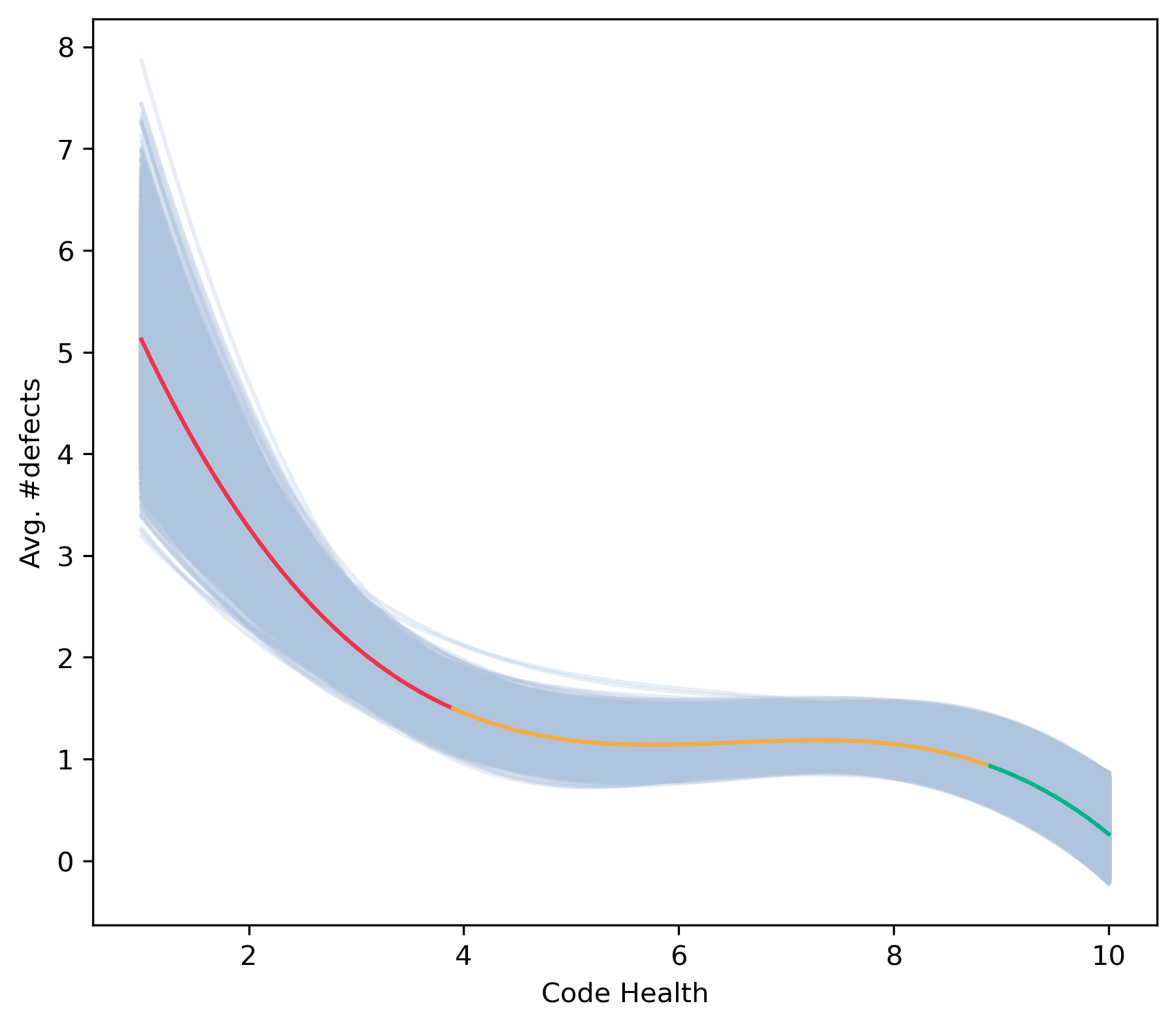}
    \caption{Average defect count per file for different CH.}
    \label{fig:regr_defects}
\end{figure}

Fig.~\ref{fig:regr_defects} shows the regression model for the average number of defects per file. The main curve delineates the central prediction, showcasing a clear decline in the average number of defects as CH increases from 1 to 5. Beyond this interval, the curve seems to plateau, indicating a stabilization in the defect counts for higher CH values. However, for the upper end of the scale, i.e., $CH \ge 8$, there is again a decrease in the number of defects.

The light blue lines surrounding the main curve show alternative averages computed using bootstrapping. We observe that the uncertainty is slightly more pronounced at lower values of CH. This resembles the distribution of our data, indicating that we should interpret the values more carefully at the lower end.

\begin{figure}
    \centering
    \includegraphics[width=0.45\textwidth]{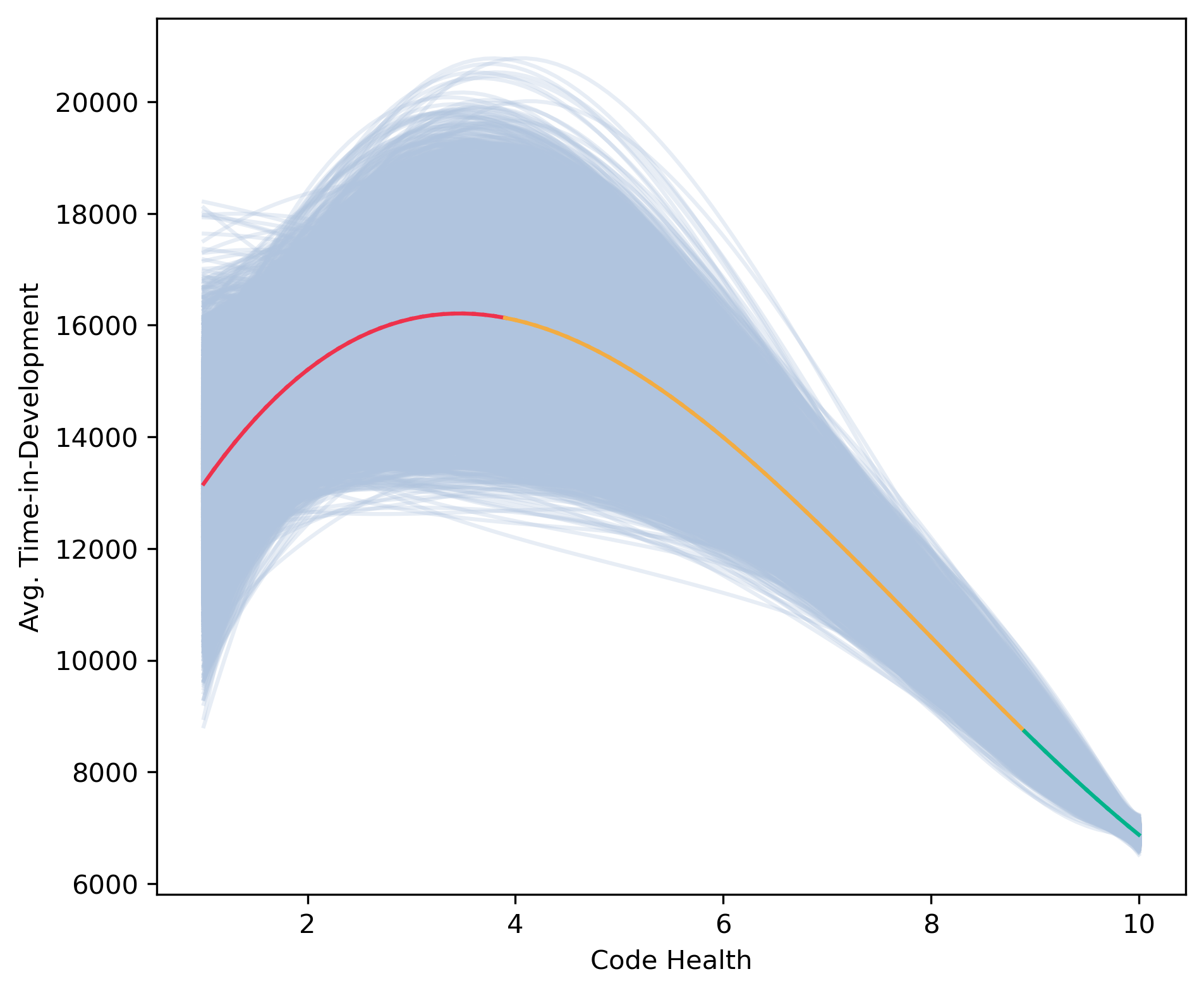}
    \caption{Average Time-in-Dev for resolving issues.}
    \label{fig:regr_time}
\end{figure}

Fig.~\ref{fig:regr_time} presents the regression model for the average Time-in-Dev when resolving Jira issues. The main curve depicts a pronounced decrease from CH 4 to 10, i.e., higher CH is associated with faster implementation of issue resolutions. The 10,000 light blue lines enveloping the main curve show individual bootstrap estimates. The dense convergence of these lines towards the higher end of the CH spectrum implies a stronger consensus in the models' predictions for higher code quality. On the other hand, the broader dispersion at lower CH values reflects higher uncertainty. It is evident that the Time-in-Dev for source code files of poor quality exhibits substantial variability, in line with findings in our original paper~\cite{tornhill_code_2022}. Notably, they allocated a distinct research question to investigate the increased uncertainty associated with resolving issues in low-quality code.

We observe an increase in the average Time-in-Dev from CH 1 to 3. As there are substantially less data representing the lower end of the CH scale, we should interpret any findings in this area with care. Still, we provide three possible explanations for the slightly counter-intuitive phenomenon observed, i.e., files with somewhat better CH might be associated with slower implementation time. 

First, the distribution of files in the lowest CH range ($CH \leq 3$) is skewed toward one. This is a result of the penalizing approach of CodeScene's CH calculation. When the number of code smells reaches very high numbers, a file reaches the bottom (CH 1). There appears to be more variation among the files with CH 1 and their average Time-in-Dev is closer to the global average in the dataset.

Two additional hypotheses revolve around the nature of software development in extremely poor code. Developers might work more swiftly yet recklessly when confronted with such code, potentially due to reduced motivation and care. If the code is cognitively demanding, they might feel less motivated to understand what is already there. Consequently, as suggested by the ``broken window'' theory~\cite{hunt_pragmatic_1999}, they might then care less about their own added solution and just get the implementation work done. 

Finally, files with CH 1 are likely to be very large. When files are very large, developers might develop ``tunnel vision'' and only focus on their small change without considering the entirety of the file. Again, this might lead to a shorter average Time-in-Dev for very low CH. Both hypotheses suggest that developers work faster and more heedlessly in very poor code, which is in line with the observation of increased bug counts for very low CH. However, future research is needed to corroborate these hypotheses.

\begin{bluebox}
a) \textbf{Defect count}: There are negative trends for $CH \leq 5$ and $CH \ge 8$. No correlation appears for $5 \leq CH \leq 8$.\\
b) \textbf{Time-in-Dev}: There is a steady decline for $CH \ge 4$ and the variability decreases. For $CH \leq 3$, but afflicted by high uncertainty, the data indicate a positive trend.\\
Observations of hasty work in files with abysmal code quality are in line with the broken windows theory.
\end{bluebox}

\subsection{Value creation (RQ2)} \label{sec:res-rq2}
This subsection presents the results from our modeling of relative changes in the value creation given different levels of CH. We investigate six different starting points representing a current CH ($CH_{0}$) and a fraction of unplanned work ($u$). We choose $CH_{0}$ representing the three CodeScene intervals \textit{alert}, \textit{warning}, and \textit{healthy}, respectively. For each $CH_{0}$, we study $u$ corresponding to a high performer and a medium performer as reported by Forsgren \textit{et al}~\cite{forsgren_accelerate_2018}.

We discuss the following starting points:
\begin{itemize}
    \item[(a)] $CH_{0} = 3.9, u=0.12$: alert CH, high-performer $u$
    \item[(b)] $CH_{0} = 3.9, u=0.25$: alert CH, medium-performer $u$  
    \item[(c)] $CH_{0} = 6.0, u=0.12$: warning CH, high-performer $u$  
    \item[(d)] $CH_{0} = 6.0, u=0.25$: warning CH, medium-performer $u$  
    \item[(e)] $CH_{0} = 9.1, u=0.12$: healthy CH, high-performer $u$  
    \item[(f)] $CH_{0} = 9.1, u=0.25$: healthy CH, Medium-performer $u$  
\end{itemize}

\begin{figure*}
        \centering
        \begin{subfigure}[b]{0.475\textwidth}
            \centering
            \includegraphics[width=\textwidth]{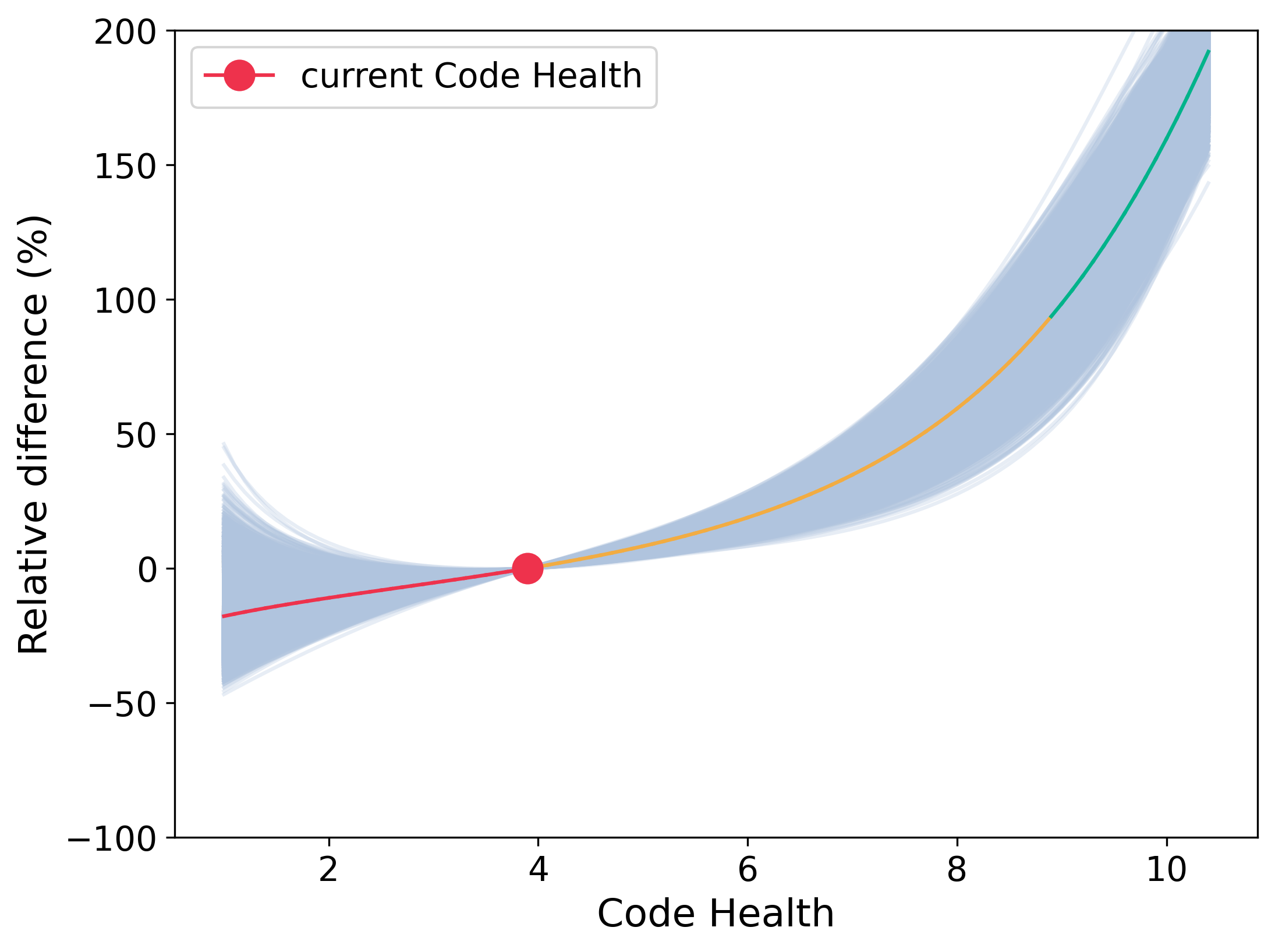}
            \caption[GreatExp A]%
            {{\small Starting point: $CH_{0} = 3.9, u=0.12$}}   
            \label{fig:value_3.9_0.12}
        \end{subfigure}
        \hfill
        \begin{subfigure}[b]{0.475\textwidth}  
            \centering 
            \includegraphics[width=\textwidth]{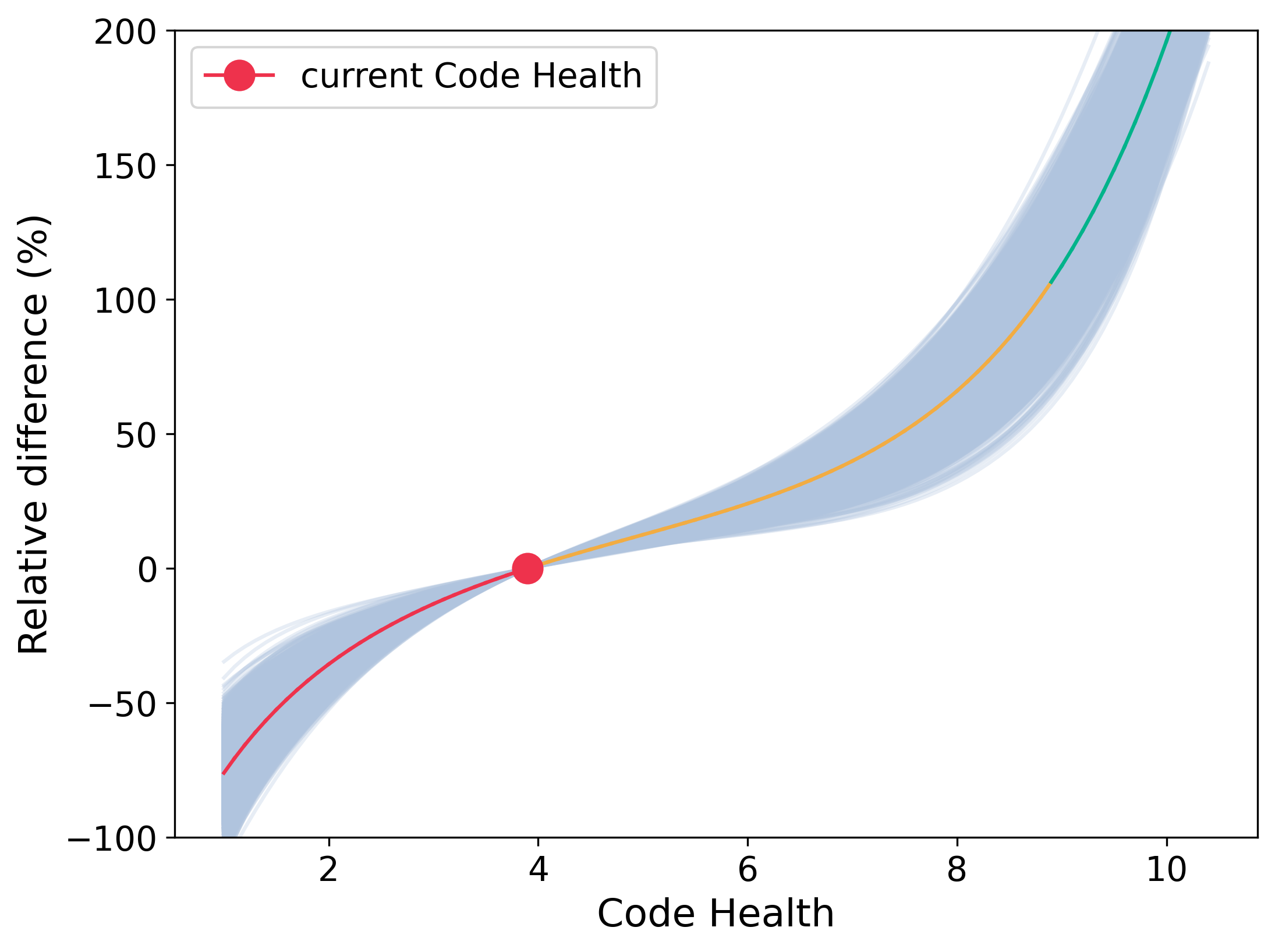}
            \caption[GreatExp B]%
            {{\small Starting point: $CH_{0} = 3.9, u=0.25$}}  
            \label{fig:value_3.9_0.25}
        \end{subfigure}
        \vskip\baselineskip
        \begin{subfigure}[b]{0.475\textwidth}   
            \centering 
            \includegraphics[width=\textwidth]{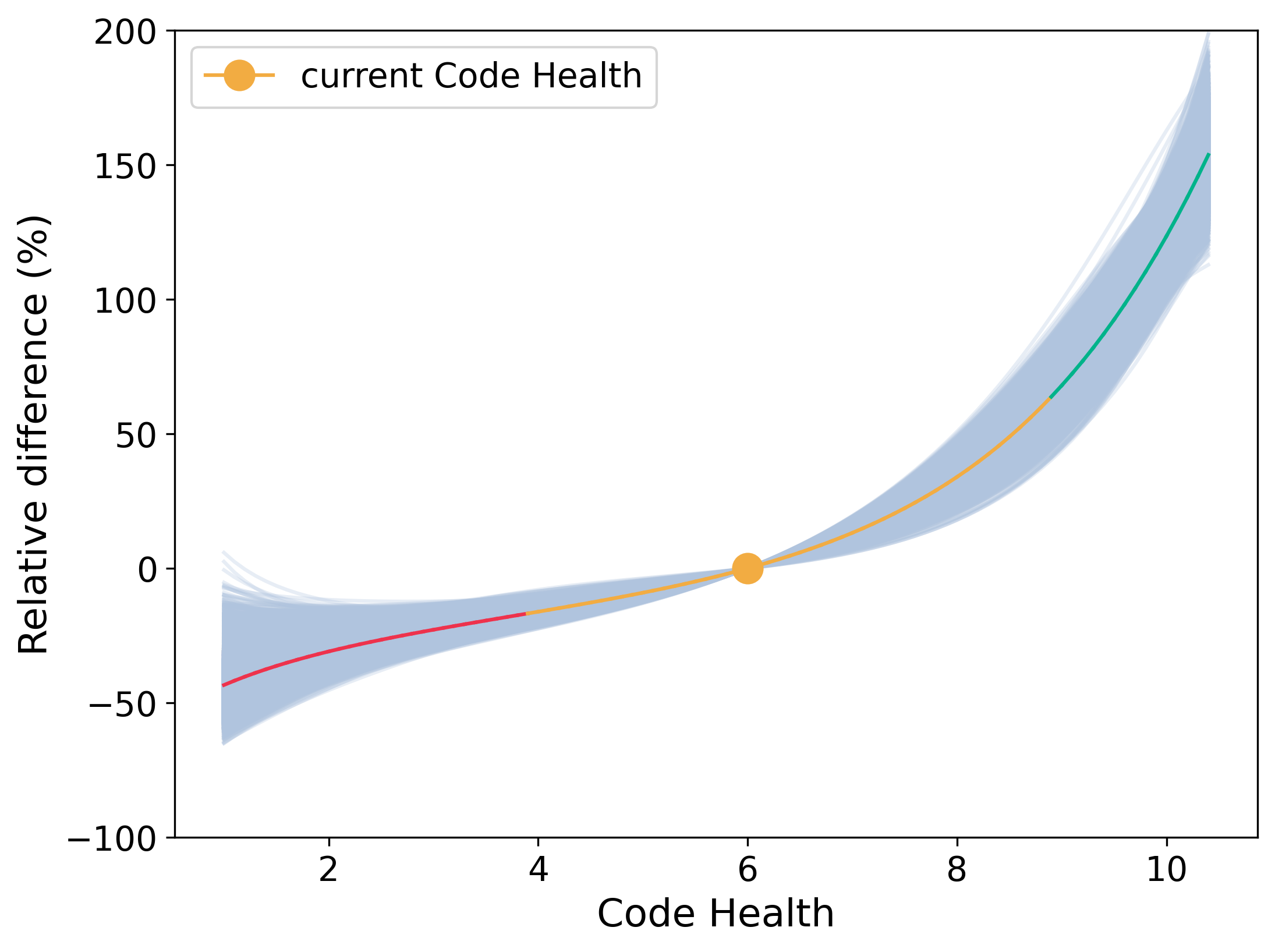}
            \caption[GreatExp C]%
            {{\small Starting point: $CH_{0} = 6.0, u=0.12$}}  
            \label{fig:value_6_0.12}
        \end{subfigure}
        \hfill
        \begin{subfigure}[b]{0.475\textwidth}   
            \centering 
            \includegraphics[width=\textwidth]{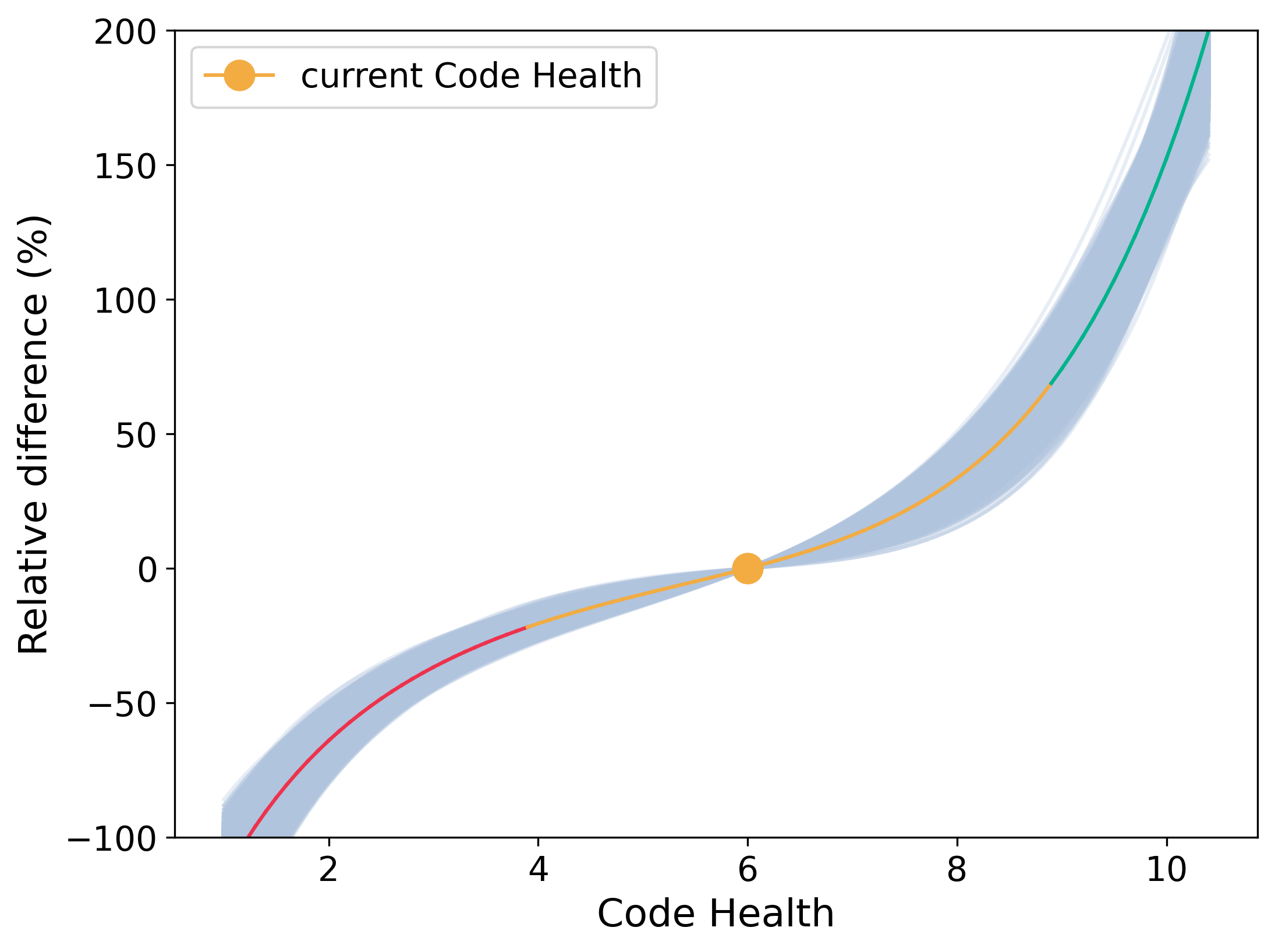}
            \caption[GreatExp D]%
            {{\small Starting point: $CH_{0} = 6.0, u=0.25$}}  
            \label{fig:value_6_0.25}
        \end{subfigure}
        \vskip\baselineskip
        \begin{subfigure}[b]{0.475\textwidth}   
            \centering 
            \includegraphics[width=\textwidth]{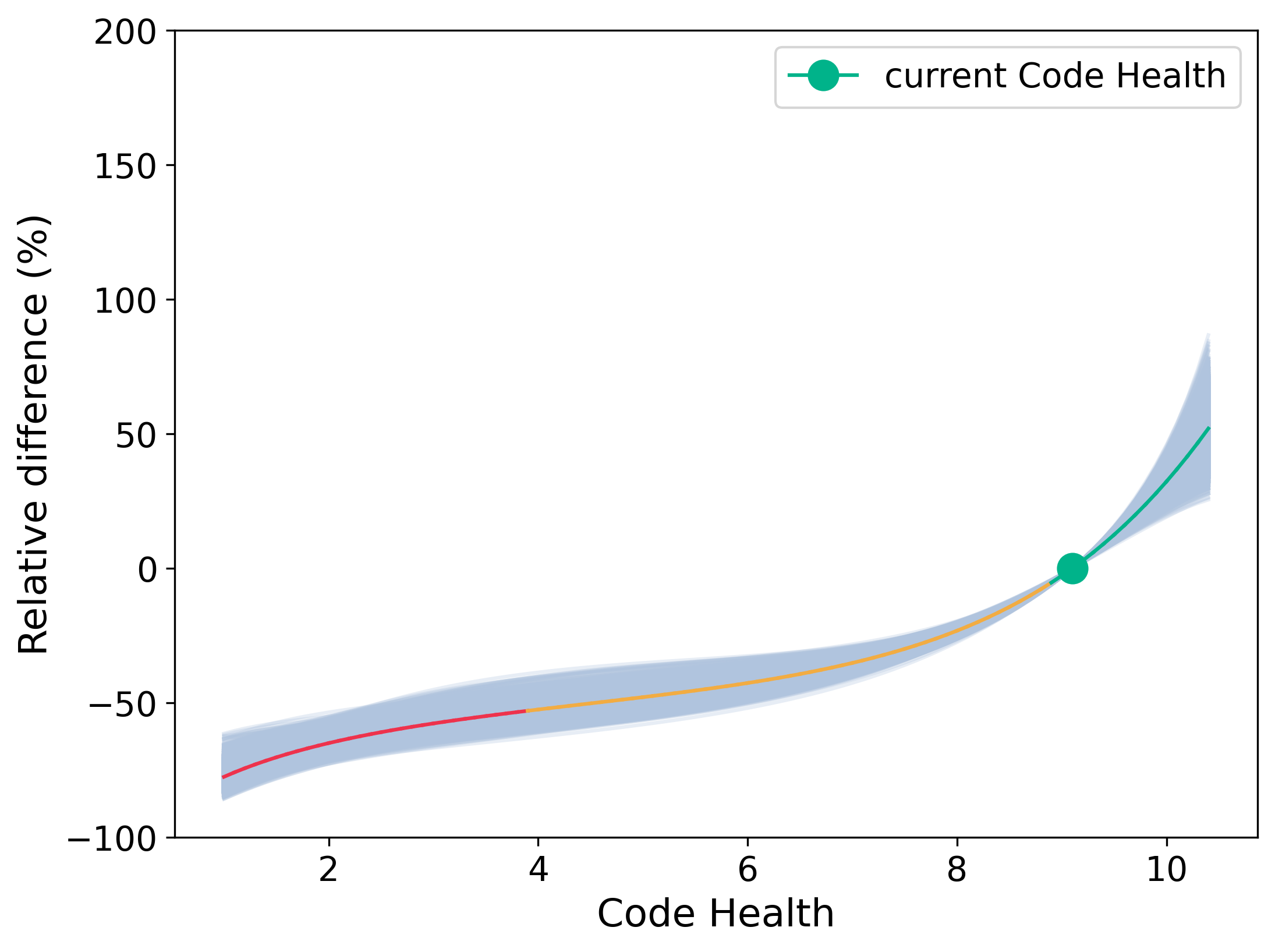}
            \caption[GreatExp C]%
            {{\small Starting point: $CH_{0} = 9.1, u=0.12$}}
            \label{fig:value_9.1_0.12}
        \end{subfigure}
        \hfill
        \begin{subfigure}[b]{0.475\textwidth}   
            \centering 
            \includegraphics[width=\textwidth]{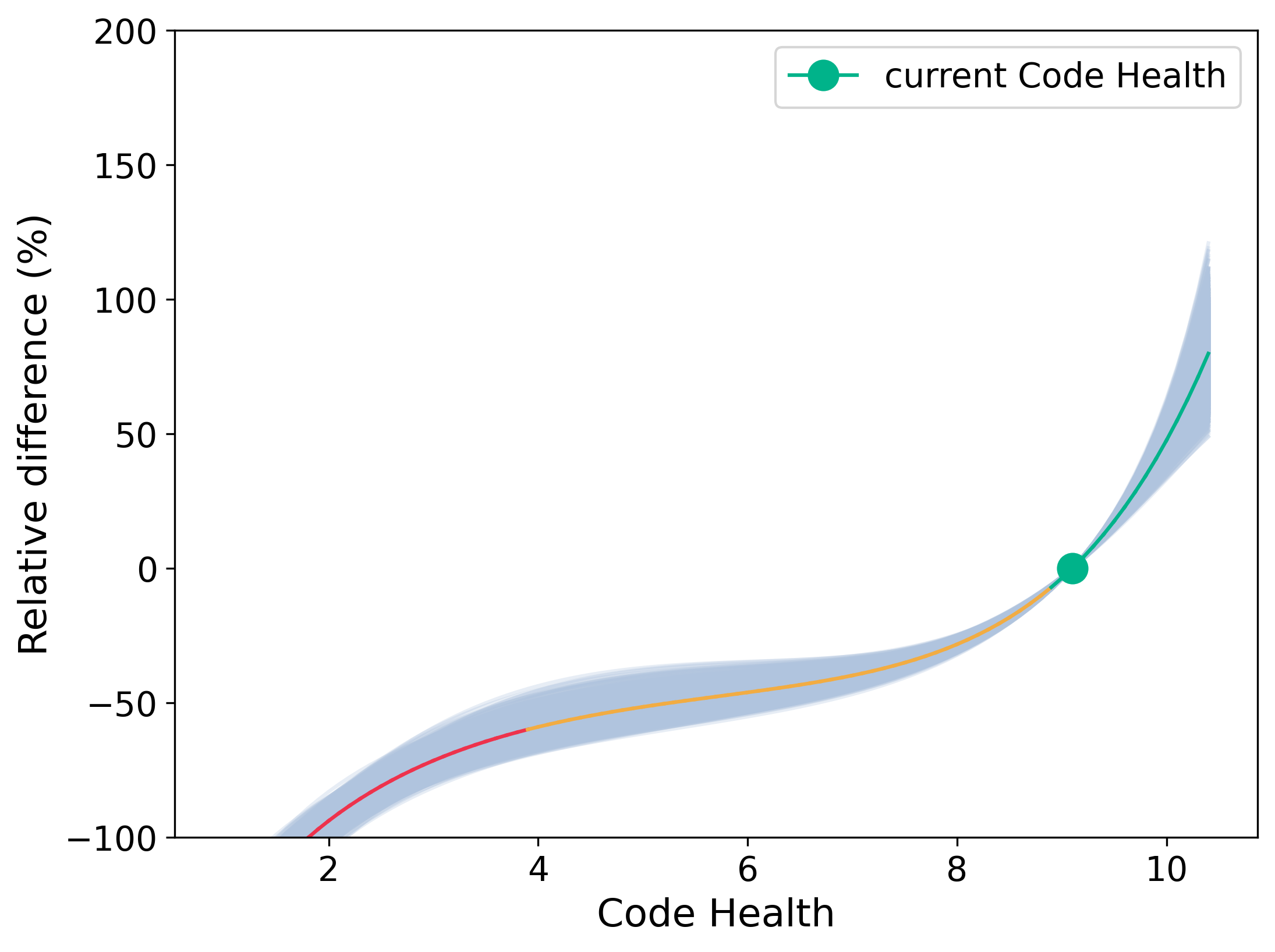}
            \caption[GreatExp D]%
            {{\small Starting point: $CH_{0} = 9.1, u=0.25$}}
            \label{fig:value_9.1_0.25}
        \end{subfigure}
        \caption[]
        {\small Relative changes of value creation for six different starting points.} 
        \label{fig:value_curves}
\end{figure*}

Fig.~\ref{fig:value_curves} presents the output from the value-creation model for six different starting points. The y-axis shows the change in value creation compared to the starting point. The main line represents value calculations based on the regression coefficients from RQ1, whereas the light blue lines display uncertainty as described in Sec.~\ref{sec:value}.

Our approach shows a clear non-linear relationship between CH and value creation. This is the most pronounced for starting points representing alert and warning CH with a medium fraction of unplanned work, i.e., Subfigures~\ref{fig:value_3.9_0.25} and~\ref{fig:value_6_0.25}. The curves display evident non-linearity at both ends. Focusing on the upper end, the curves steepens sharply from $CH=8$. This highlights an amplified sensitivity to changes at higher CH values. For these two starting points ((b) and (d)), changes to CH in the interval from 3 to 7 are less prominent. For very low levels of Code Health, however, the increased curvature shows another example of non-linearity. Subfigure~\ref{fig:value_9.1_0.25}, representing healthy CH and a medium fraction of unplanned work, depicts similar tendencies.

Subfigures~\ref{fig:value_3.9_0.12},~\ref{fig:value_6_0.12}, and~\ref{fig:value_9.1_0.12} represent contexts with a low fraction of unplanned work ($u=0.12$) --- corresponding to high performers. Analogously to the medium performers, the curves depicting alert- and warning levels of CH steepen sharply from CH 8. This indicates that also for high performers, there remains substantial potential value in increasing the code quality to the highest levels. 

On the other hand, Subfigures~\ref{fig:value_3.9_0.12},~\ref{fig:value_6_0.12}, and~\ref{fig:value_9.1_0.12} display less evident non-linearity for CH below the starting points. This finding suggests that high performers with very low levels of unplanned work have a more robust value creation process, i.e., they might experience less dramatic effects if CH decreases.

All subfigures in Fig.~\ref{fig:value_curves} illustrate the uncertainty of the average values. The light blue lines, resulting in shaded regions flanking the main curves, represent alternative value functions sampled from the regression models. As expected, we find a greater variability further away from the starting points. Furthermore, we notice that the value functions representing lower CH (Subfigures~\ref{fig:value_3.9_0.12} and\ref{fig:value_3.9_0.25}) involve more uncertainty than the corresponding functions for high CH (cf. Subfigures~\ref{fig:value_9.1_0.12} and\ref{fig:value_9.1_0.25}). This can be explained by the data distribution, i.e., only a small fraction of the dataset represents alert-level CH.

\begin{bluebox}
The value model illustrates strong non-linearities at the ends of the CH interval. For the upper end, i.e., $CH \ge 8$, the plot indicates increasing --- not diminishing --- returns of value.
\end{bluebox}

Our value creation model is based on file-level regression models. To initiate discussions on the importance of highly maintainable source code, we will now speculate about to what extent this can be extrapolated to entire projects --- an important avenue for future work. We discuss the possibility of exploring differences for 1) averages of all files and 2) averages among the most active files. 

First, one could initiate quality discussion on the project level by investigating average CH scores. This metric is straightforward to interpret but could be overly simplistic, leading to potential biases. For instance, a project with numerous small, high-quality files would display an inflated average CH. However, this could be mitigated by calculating a weighted average based on file size, i.e., larger files would then influence the average more. However, discussing the potential of increasing the quality of the average file, would still overstate expectations for files with no activity. There is little value in paying down TD in such files with a low interest rate. 

Second, one could instead focus discussions on average CH for files that frequently change. These files contain code that developers frequently must comprehend, modify, and extend --- files for which the interest rate on the TD is high. Again, weighting the average based on file size would be needed to avoid bias of small files. A weighted CH average of files that at a given point in time have a relatively high level of churn would likely be a better option. We plan to explore this pragmatic approach as a conversation starter in collaboration with industry partners in future work. While a crude approach to return on investment discussions for refactoring efforts, the initial feedback from pilot runs has been positive. 

\begin{bluebox}
Extrapolating file-level results to project-wide averages is an important research direction. Weighted averages of churn-filtered files can trigger useful discussions related to return on investment in refactoring efforts. 
\end{bluebox}

\section{Limitations of the value-creation model} \label{sec:limit}
As a very simple model, it is obvious that our value-creation model suffers from limitations. However, as George Box famously put it ``\textit{All models are wrong, but some are useful}.’’ The question is whether our model can be put to use. In this section, we elaborate on five major limitations of the model.

\textbf{Extrapolating from the file level. } Sec.~\ref{sec:res-rq2} speculates about how the value-creation model could be discussed on a project level. Our suggestion is to use \emph{churn-based filtering} and \emph{size-based weighting} to provide averages that can be discussed with project stakeholders. Nevertheless, this is a reductionistic bottom-up approach that can never capture the complete picture of software engineering complexities. On the other hand, discussion based on the model should remain at the highest level to only discuss relative changes from a baseline. We argue that this is an accessible approach that can help engineering managers communicate the business value of source code quality to upper management and board room discussions.

\textbf{File size is not controlled for.} While we discuss approaches to use filtering and weighting of files, this has not yet been part of our analysis. Regarding our regression models, we strongly believe that we could capture more of the data variability if we controlled for file size, i.e., Lines of Code (LoC). Both file size and churn are known to be predictors of defect-prone files~\cite{koru_investigation_2009,giger_comparing_2011}. Unfortunately, neither of these aspects is represented in the merged datasets we study. However, in our original study, we investigated file size and found that Code Health has a higher degree of correlation than LoC. In future work, we plan to collect a dataset that contains both size and churn.

\textbf{Coupling of files is not considered.} Source code files in a repository are anything but independent, which means an underlying assumption in regression modeling is violated. The observations in the dataset are not independent of each other, thus the CH value of one file might be correlated with another. Our merged dataset does not allow us to measure the extent of change coupling. However, commits tend to touch several files at once. Unfortunately, this violation is common in software engineering research --- additional future work is needed to tackle this issue, along the lines of work by Kitchenham and colleagues~\cite{kitchenham_robust_2017}.

\textbf{Capacity is not only about Time-in-Dev.} Assuming an inverse proportionality between Time-in-Dev and capacity is an oversimplification. Software engineering productivity is an entire research area~\cite{forsgren_space_2021} and our proposed construct \textit{capacity} should ideally be treated in a similarly multifaceted way. We acknowledge that simply implementing code related to Jira issues faster provides a narrow-minded view on capacity increases --- most importantly, this approach disregards the quality of the issue resolutions. Quick and dirty fixes would inflate capacity using our model.

\textbf{Unplanned work is not only about defects.} While there is a clear relationship between defects and unplanned work, it only explains a part of the picture. Not all unplanned work arises from defects. Other factors, such as changes in business priorities, external dependencies, emergencies, or even new opportunities, can also contribute to unplanned work. We simplistically assume a direct proportional relationship. In reality, a few critical defects can cause more unplanned work than many minor ones.

\section{Threats to Validity} \label{sec:threats}
Our study inherits threats to validity from previously collected datasets, and this section explores both these and additional new threats.

\textbf{External validity} refers to the extent to which the findings can be generalized to software engineering in general. The dataset is large and representative of contemporary software engineering based on the variety of programming languages, companies, and industry segments being represented. Still, all included data originate in companies that are CodeScene customers, which might bias the results toward organizations that are more inclined to pay for advanced software tools. It is likely that very small companies and startups are not represented in the dataset. In contrast to most academic work, there is no code from open-source software in this dataset. However, based on the many languages covered, and a mix of back-end and front-end code, we believe our findings properly represent current industry practice.

\textbf{Construct validity} relates to whether our introduced concepts are appropriate and to what degree our measures effectively represent them~\cite{sjoberg_improving_2023}. Our work discusses the four constructs 1) \emph{code quality}, 2) \emph{defect count}, 3) \emph{Time-in-Dev}, and 4) \emph{value creation}. As discussed in Sec.~\ref{sec:bg}, quality is always a multi-faceted concept that is hard to pinpoint. As CH targets the maintainability dimension, the whole picture of quality is not captured --- but it is an appropriate measure for our TD perspective. The second one is an established construct that we measure using commits connected to Jira issues tagged ``bug.'' As discussed by Herzig \textit{et al.}~\cite{herzig_its_2013}, many issues in open-source software projects might be mislabeled. We cannot assess the fraction of mislabelled issues in the CodeRed and UOwns datasets. We suspect that the label correctness is generally higher in proprietary projects, but we cannot know. 

Construct~3 has been discussed in independent research~\cite{tawosi_relationship_2022}. As discussed in the original papers, how carefully developers change the status of an assigned Jira issue relies on discipline~\cite{tornhill_code_2022}. Thus, small variations (e.g., context switches and interruptions) will not be captured~\cite{borg_u_2023}. Our mitigation strategy is to rely on a large dataset to let averaging effects support the robustness of our measurements. The construct ``value creation'' represents a bigger threat --- this is a novel proposal to describe an abstract business phenomenon. We acknowledge that our multiplication of the two factors $capacity$ and $(1 - \%unplannedWork)$ remains on an abstract level. Moreover, capacity and unplanned work are constructs in their own right. However, we accept this threat as we only use the measures to discuss relative differences.

\textbf{Internal validity} is crucial for establishing causality. Our study, however, is based on data from two observational studies, which makes causal claims inherently subject to scrutiny. 

In RQ1, we speculate about the relationship between the dependent variable CH on the one hand, and defect count and Time-in-Dev on the other hand. Software engineering is an intricate activity with numerous confounding factors at play. File size and change coupling have been discussed in Sec.~\ref{sec:limit}. Other categories of confounders include organizational, process, business, and people aspects. We mitigate these threats by combining data from many development contexts and by limiting our discussion to general patterns. Moreover, we measure file-level CH at the time of the most recent commit during the 1-year window (see Sec.~\ref{sec:data}), which means we assume it has remained stable during the period. While CH might have fluctuated over several commits for some files, we mitigate this threat by analyzing a reasonably short time window. A longer window could have yielded different results.

In RQ2, we discuss how CH is associated with value creation. According to causality pioneer Judea Pearl’s ``Ladder of Causality’’~\cite{pearl_book_2018} there are three rungs: 1) association (correlations and patterns), 2) intervention (effects of actions), and 3) counterfactuals (hypotheticals and alternative realities). Our observational data ground us on the first rung. While the causal analysis performed by Léven \emph{et al.} has taken TD research to the second rung, we should not disregard rung-1 studies. As discussed by Giancolo \emph{et al.} in the context of epidemiology~\cite{gianicolo_methods_2020}, causality is often probabilistic in nature. Any epidemiological (or medical) intervention can typically only alter the probability that a desired event will take place. We argue that a similar viewpoint is valuable in the software engineering context. High CH is likely to increase the probability that a source code file will contain fewer defects and be faster to maintain. Both these advantages will likely support an organization’s value creation.

\textbf{Conclusion validity} is the degree to which conclusions drawn about the relationships observed in the data are reasonable and appropriate. We focus the discussion on our use of statistics. Despite the low $R^{2}$ scores indicating a poor goodness-of-fit in our RQ1 regression models, they remain insightful for identifying overarching patterns. Regarding defect count, the large dataset shows stable average numbers. For Time-in-Dev, we observe significant individual file-level variations that deserve future work based on more metadata. By controlling for key confounders, we could provide a more nuanced understanding.

For RQ2, we present plots originating from six different starting points to illustrate our value-creation model. Our model simplifies the relationships between variables by assuming direct proportional changes. To avoid potential misinterpretations, we focus our discussion on the general curvature trends across different CH intervals. The inclusion of uncertainty bands alongside the six starting points serves as a sensitivity analysis. Nevertheless, we view these results primarily as a conversation starter, deliberately refraining from quantifying explicit value changes.

\section{Conclusion} \label{sec:conc}
We build on previous work by presenting a more fine-grained analysis of the associations between Code Health on the one hand, and defect count and Time-in-Dev on the other hand. Moreover, we propose a simple value-creation model to spark discussion on relative changes in business value.

Our results suggest that improving code maintainability from very high to excellent pays off. The returns in the upper end of the quality spectrum are not diminishing --- they are increasing. This underscores that exceptional code quality is not only a developer's vanity metric. Technical roles must better communicate the benefits to executive levels to inform strategic refactoring decisions.

Furthermore, our findings motivate a proactive ``zero-tolerance'' policy toward code smells in files where the interest rate on the technical debt is high, i.e., files with substantial churn. Thus, our work supports the ideas outlined in the ``broken windows'' theory~\cite{hunt_pragmatic_1999}. On the lower end of the quality interval, our findings are also in line with the theory, i.e., developers appear to resolve issues fast --- and possibly recklessly --- if the code quality is already abysmal.

Our study serves as a stepping-stone toward further research on evidence-based technical debt management. The immediate priority is to acquire more comprehensive data, which is crucial for controlling key confounding factors. Subsequently, we aim to contribute to the growing body of evidence by performing causal analyses. Additionally, future research should focus on integrating empirical findings into decision-support frameworks for conducting debt trade-off analyses. Although various frameworks have been proposed~\cite{lenarduzzi_systematic_2021}, research is needed to create a practical, user-friendly tool suitable for industry use.

\section*{Data Availability}
The replication package is available on Zenodo~\cite{notebook}.

\balance
\bibliographystyle{ACM-Reference-Format}
\bibliography{codered}


\begin{thebibliography}{43}


\ifx \showCODEN    \undefined \def \showCODEN     #1{\unskip}     \fi
\ifx \showDOI      \undefined \def \showDOI       #1{#1}\fi
\ifx \showISBNx    \undefined \def \showISBNx     #1{\unskip}     \fi
\ifx \showISBNxiii \undefined \def \showISBNxiii  #1{\unskip}     \fi
\ifx \showISSN     \undefined \def \showISSN      #1{\unskip}     \fi
\ifx \showLCCN     \undefined \def \showLCCN      #1{\unskip}     \fi
\ifx \shownote     \undefined \def \shownote      #1{#1}          \fi
\ifx \showarticletitle \undefined \def \showarticletitle #1{#1}   \fi
\ifx \showURL      \undefined \def \showURL       {\relax}        \fi
\providecommand\bibfield[2]{#2}
\providecommand\bibinfo[2]{#2}
\providecommand\natexlab[1]{#1}
\providecommand\showeprint[2][]{arXiv:#2}

\bibitem[\protect\citeauthoryear{Ayala, Turhan, Franch, and Juristo}{Ayala
  et~al\mbox{.}}{2022}]%
        {ayala_use_2022}
\bibfield{author}{\bibinfo{person}{Claudia Ayala}, \bibinfo{person}{Burak
  Turhan}, \bibinfo{person}{Xavier Franch}, {and} \bibinfo{person}{Natalia
  Juristo}.} \bibinfo{year}{2022}\natexlab{}.
\newblock \showarticletitle{Use and {Misuse} of the {Term} “{Experiment}”
  in {Mining} {Software} {Repositories} {Research}}.
\newblock \bibinfo{journal}{\emph{IEEE Trans. on Software Engineering}}
  \bibinfo{volume}{48}, \bibinfo{number}{11} (\bibinfo{year}{2022}),
  \bibinfo{pages}{4229--4248}.
\newblock
\showISSN{1939-3520}


\bibitem[\protect\citeauthoryear{Baggen, Correia, Schill, and Visser}{Baggen
  et~al\mbox{.}}{2012}]%
        {baggen_standardized_2012}
\bibfield{author}{\bibinfo{person}{Robert Baggen}, \bibinfo{person}{José~Pedro
  Correia}, \bibinfo{person}{Katrin Schill}, {and} \bibinfo{person}{Joost
  Visser}.} \bibinfo{year}{2012}\natexlab{}.
\newblock \showarticletitle{Standardized {Code} {Quality} {Benchmarking} for
  {Improving} {Software} {Maintainability}}.
\newblock \bibinfo{journal}{\emph{Software Quality Journal}}
  \bibinfo{number}{2} (\bibinfo{year}{2012}), \bibinfo{pages}{287--307}.
\newblock


\bibitem[\protect\citeauthoryear{Borg, Pruvost, Mones, and Tornhill}{Borg
  et~al\mbox{.}}{2024}]%
        {notebook}
\bibfield{author}{\bibinfo{person}{Markus Borg}, \bibinfo{person}{Ilyana
  Pruvost}, \bibinfo{person}{Enys Mones}, {and} \bibinfo{person}{Adam
  Tornhill}.} \bibinfo{year}{2024}\natexlab{}.
\newblock \bibinfo{title}{Accompanying Replication Package}.
\newblock
\newblock
\urldef\tempurl%
\url{https://zenodo.org/records/10560722}
\showURL{%
\tempurl}


\bibitem[\protect\citeauthoryear{Borg, Tornhill, and Mones}{Borg
  et~al\mbox{.}}{2023}]%
        {borg_u_2023}
\bibfield{author}{\bibinfo{person}{Markus Borg}, \bibinfo{person}{Adam
  Tornhill}, {and} \bibinfo{person}{Enys Mones}.}
  \bibinfo{year}{2023}\natexlab{}.
\newblock \showarticletitle{U {Owns} the {Code} {That} {Changes} and {How}
  {Marginal} {Owners} {Resolve} {Issues} {Slower} in {Low}-{Quality} {Source}
  {Code}}. In \bibinfo{booktitle}{\emph{Proc. of the 27th {Int'l.} {Conf.} on
  {Evaluation} and {Assessment} in {Software} {Engineering}}}.
  \bibinfo{pages}{368--377}.
\newblock
\showISBNx{9798400700446}
\urldef\tempurl%
\url{https://doi.org/10.1145/3593434.3593480}
\showDOI{\tempurl}


\bibitem[\protect\citeauthoryear{Chatterjee, Sharma, and Ralph}{Chatterjee
  et~al\mbox{.}}{2022}]%
        {chatterjee_empirical_2022}
\bibfield{author}{\bibinfo{person}{Preetha Chatterjee}, \bibinfo{person}{Tushar
  Sharma}, {and} \bibinfo{person}{Paul Ralph}.}
  \bibinfo{year}{2022}\natexlab{}.
\newblock \showarticletitle{Empirical {Standards} for {Repository} {Mining}}.
  In \bibinfo{booktitle}{\emph{Proc. of the 19th {Int'l.} {Conf.} on {Mining}
  {Software} {Repositories}}}. \bibinfo{pages}{142--143}.
\newblock
\showISBNx{978-1-4503-9303-4}
\urldef\tempurl%
\url{https://doi.org/10.1145/3524842.3528032}
\showDOI{\tempurl}


\bibitem[\protect\citeauthoryear{Efron}{Efron}{1979}]%
        {efron_bootstrap_1979}
\bibfield{author}{\bibinfo{person}{B. Efron}.} \bibinfo{year}{1979}\natexlab{}.
\newblock \showarticletitle{Bootstrap {Methods}: {Another} {Look} at the
  {Jackknife}}.
\newblock \bibinfo{journal}{\emph{The Annals of Statistics}}
  \bibinfo{volume}{7}, \bibinfo{number}{1} (\bibinfo{year}{1979}),
  \bibinfo{pages}{1--26}.
\newblock
\showISSN{0090-5364, 2168-8966}
\urldef\tempurl%
\url{https://doi.org/10.1214/aos/1176344552}
\showDOI{\tempurl}


\bibitem[\protect\citeauthoryear{Farley}{Farley}{2021}]%
        {farley_modern_2021}
\bibfield{author}{\bibinfo{person}{David Farley}.}
  \bibinfo{year}{2021}\natexlab{}.
\newblock \bibinfo{booktitle}{\emph{Modern {Software} {Engineering}: {Doing}
  {What} {Works} to {Build} {Better} {Software} {Faster}}}.
\newblock \bibinfo{publisher}{Addison-Wesley Professional},
  \bibinfo{address}{Boston, MA, USA}.
\newblock


\bibitem[\protect\citeauthoryear{Fenton}{Fenton}{1994}]%
        {fenton_software_1994}
\bibfield{author}{\bibinfo{person}{Norman Fenton}.}
  \bibinfo{year}{1994}\natexlab{}.
\newblock \showarticletitle{Software {Measurement}: {A} {Necessary}
  {Scientific} {Basis}}.
\newblock \bibinfo{journal}{\emph{IEEE Trans. on Software Engineering}}
  \bibinfo{volume}{20}, \bibinfo{number}{3} (\bibinfo{year}{1994}),
  \bibinfo{pages}{199--206}.
\newblock


\bibitem[\protect\citeauthoryear{Forsgren, Humble, and Kim}{Forsgren
  et~al\mbox{.}}{2018}]%
        {forsgren_accelerate_2018}
\bibfield{author}{\bibinfo{person}{Nicole Forsgren}, \bibinfo{person}{Jez
  Humble}, {and} \bibinfo{person}{Gene Kim}.} \bibinfo{year}{2018}\natexlab{}.
\newblock \bibinfo{booktitle}{\emph{Accelerate: {The} {Science} of {Lean}
  {Software} and {DevOps}: {Building} and {Scaling} {High} {Performing}
  {Technology} {Organizations}} (\bibinfo{edition}{1} ed.)}.
\newblock \bibinfo{publisher}{IT Revolution Press}, \bibinfo{address}{Portland,
  Oregon}.
\newblock
\showISBNx{978-1-942788-33-1}


\bibitem[\protect\citeauthoryear{Forsgren, Storey, Maddila, Zimmermann, Houck,
  and Butler}{Forsgren et~al\mbox{.}}{2021}]%
        {forsgren_space_2021}
\bibfield{author}{\bibinfo{person}{Nicole Forsgren},
  \bibinfo{person}{Margaret-Anne Storey}, \bibinfo{person}{Chandra Maddila},
  \bibinfo{person}{Thomas Zimmermann}, \bibinfo{person}{Brian Houck}, {and}
  \bibinfo{person}{Jenna Butler}.} \bibinfo{year}{2021}\natexlab{}.
\newblock \showarticletitle{The {SPACE} of {Developer} {Productivity}:
  {There}'s more to it than you think.}
\newblock \bibinfo{journal}{\emph{Queue}} \bibinfo{volume}{19},
  \bibinfo{number}{1} (\bibinfo{year}{2021}), \bibinfo{pages}{10:20--10:48}.
\newblock
\showISSN{1542-7730}
\urldef\tempurl%
\url{https://doi.org/10.1145/3454122.3454124}
\showDOI{\tempurl}


\bibitem[\protect\citeauthoryear{Foucault, Falleri, and Blanc}{Foucault
  et~al\mbox{.}}{2014}]%
        {foucault_code_2014}
\bibfield{author}{\bibinfo{person}{Matthieu Foucault},
  \bibinfo{person}{Jean-Rémy Falleri}, {and} \bibinfo{person}{Xavier Blanc}.}
  \bibinfo{year}{2014}\natexlab{}.
\newblock \showarticletitle{Code {Ownership} in {Open}-{Source} {Software}}. In
  \bibinfo{booktitle}{\emph{Proc. of the 18th {Int'l.} {Conf.} on {Evaluation}
  and {Assessment} in {Software} {Engineering}}}. \bibinfo{pages}{1--9}.
\newblock
\showISBNx{978-1-4503-2476-2}
\urldef\tempurl%
\url{https://doi.org/10.1145/2601248.2601283}
\showURL{%
\tempurl}


\bibitem[\protect\citeauthoryear{Gianicolo, Eichler, Muensterer, Strauch, and
  Blettner}{Gianicolo et~al\mbox{.}}{2020}]%
        {gianicolo_methods_2020}
\bibfield{author}{\bibinfo{person}{Emilio Gianicolo}, \bibinfo{person}{Martin
  Eichler}, \bibinfo{person}{Oliver Muensterer}, \bibinfo{person}{Konstantin
  Strauch}, {and} \bibinfo{person}{Maria Blettner}.}
  \bibinfo{year}{2020}\natexlab{}.
\newblock \showarticletitle{Methods for {Evaluating} {Causality} in
  {Observational} {Studies}}.
\newblock \bibinfo{journal}{\emph{Deutsches Arzteblatt Int'l.}}
  \bibinfo{volume}{117}, \bibinfo{number}{7} (\bibinfo{year}{2020}),
  \bibinfo{pages}{101--107}.
\newblock
\showISSN{1866-0452}
\urldef\tempurl%
\url{https://doi.org/10.3238/arztebl.2020.0101}
\showDOI{\tempurl}


\bibitem[\protect\citeauthoryear{Giger, Pinzger, and Gall}{Giger
  et~al\mbox{.}}{2011}]%
        {giger_comparing_2011}
\bibfield{author}{\bibinfo{person}{Emanuel Giger}, \bibinfo{person}{Martin
  Pinzger}, {and} \bibinfo{person}{Harald Gall}.}
  \bibinfo{year}{2011}\natexlab{}.
\newblock \showarticletitle{Comparing {Fine}-grained {Source} {Code} {Changes}
  and {Code} {Churn} for {Bug} {Prediction}}. In
  \bibinfo{booktitle}{\emph{Proc. of the 8th {Working} {Conf.} on {Mining}
  {Software} {Repositories}}}. \bibinfo{pages}{83--92}.
\newblock
\urldef\tempurl%
\url{https://dl.acm.org/doi/abs/10.1145/1985441.1985456}
\showURL{%
\tempurl}


\bibitem[\protect\citeauthoryear{Hall, Beecham, Bowes, Gray, and Counsell}{Hall
  et~al\mbox{.}}{2012}]%
        {hall_systematic_2012}
\bibfield{author}{\bibinfo{person}{Tracy Hall}, \bibinfo{person}{Sarah
  Beecham}, \bibinfo{person}{David Bowes}, \bibinfo{person}{David Gray}, {and}
  \bibinfo{person}{Steve Counsell}.} \bibinfo{year}{2012}\natexlab{}.
\newblock \showarticletitle{A {Systematic} {Literature} {Review} on {Fault}
  {Prediction} {Performance} in {Software} {Engineering}}.
\newblock \bibinfo{journal}{\emph{IEEE Trans. on Software Engineering}}
  \bibinfo{volume}{38}, \bibinfo{number}{6} (\bibinfo{year}{2012}),
  \bibinfo{pages}{1276--1304}.
\newblock
\showISSN{1939-3520}
\urldef\tempurl%
\url{https://doi.org/10.1109/TSE.2011.103}
\showDOI{\tempurl}


\bibitem[\protect\citeauthoryear{Herzig, Just, and Zeller}{Herzig
  et~al\mbox{.}}{2013}]%
        {herzig_its_2013}
\bibfield{author}{\bibinfo{person}{Kim Herzig}, \bibinfo{person}{Sascha Just},
  {and} \bibinfo{person}{Andreas Zeller}.} \bibinfo{year}{2013}\natexlab{}.
\newblock \showarticletitle{It's {Not} a {Bug}, {It}'s a {Feature}: {How}
  {Misclassification} {Impacts} {Bug} {Prediction}}. In
  \bibinfo{booktitle}{\emph{Proc. of the 35th {Int'l.} {Conf.} on {Software}
  {Engineering}}}. \bibinfo{pages}{392--401}.
\newblock
\urldef\tempurl%
\url{https://doi.org/10.1109/ICSE.2013.6606585}
\showDOI{\tempurl}


\bibitem[\protect\citeauthoryear{Hosseini, Turhan, and Gunarathna}{Hosseini
  et~al\mbox{.}}{2019}]%
        {hosseini_systematic_2019}
\bibfield{author}{\bibinfo{person}{Seyedrebvar Hosseini},
  \bibinfo{person}{Burak Turhan}, {and} \bibinfo{person}{Dimuthu Gunarathna}.}
  \bibinfo{year}{2019}\natexlab{}.
\newblock \showarticletitle{A {Systematic} {Literature} {Review} and
  {Meta}-{Analysis} on {Cross} {Project} {Defect} {Prediction}}.
\newblock \bibinfo{journal}{\emph{IEEE Trans. on Software Engineering}}
  \bibinfo{volume}{45}, \bibinfo{number}{2} (\bibinfo{year}{2019}),
  \bibinfo{pages}{111--147}.
\newblock
\showISSN{1939-3520}
\urldef\tempurl%
\url{https://doi.org/10.1109/TSE.2017.2770124}
\showDOI{\tempurl}


\bibitem[\protect\citeauthoryear{Hunt and Thomas}{Hunt and Thomas}{1999}]%
        {hunt_pragmatic_1999}
\bibfield{author}{\bibinfo{person}{Andrew Hunt} {and} \bibinfo{person}{David
  Thomas}.} \bibinfo{year}{1999}\natexlab{}.
\newblock \bibinfo{booktitle}{\emph{The {Pragmatic} {Programmer}: {From}
  {Journeyman} to {Master}} (\bibinfo{edition}{1} ed.)}.
\newblock \bibinfo{publisher}{Addison-Wesley Professional},
  \bibinfo{address}{Boston, MA, USA}.
\newblock
\showISBNx{978-0-201-61622-4}


\bibitem[\protect\citeauthoryear{Junior and Travassos}{Junior and
  Travassos}{2022}]%
        {junior_consolidating_2022}
\bibfield{author}{\bibinfo{person}{Helvio~Jeronimo Junior} {and}
  \bibinfo{person}{Guilherme~Horta Travassos}.}
  \bibinfo{year}{2022}\natexlab{}.
\newblock \showarticletitle{Consolidating a {Common} {Perspective} on
  {Technical} {Debt} and its {Management} {Through} a {Tertiary} {Study}}.
\newblock \bibinfo{journal}{\emph{Information and Software Technology}}
  \bibinfo{volume}{149} (\bibinfo{year}{2022}), \bibinfo{pages}{106964}.
\newblock
\showISSN{0950-5849}


\bibitem[\protect\citeauthoryear{Kitchenham, Madeyski, Budgen, Keung, Brereton,
  Charters, Gibbs, and Pohthong}{Kitchenham et~al\mbox{.}}{2017}]%
        {kitchenham_robust_2017}
\bibfield{author}{\bibinfo{person}{Barbara Kitchenham}, \bibinfo{person}{Lech
  Madeyski}, \bibinfo{person}{David Budgen}, \bibinfo{person}{Jacky Keung},
  \bibinfo{person}{Pearl Brereton}, \bibinfo{person}{Stuart Charters},
  \bibinfo{person}{Shirley Gibbs}, {and} \bibinfo{person}{Amnart Pohthong}.}
  \bibinfo{year}{2017}\natexlab{}.
\newblock \showarticletitle{Robust {Statistical} {Methods} for {Empirical}
  {Software} {Engineering}}.
\newblock \bibinfo{journal}{\emph{Empirical Software Engineering}}
  \bibinfo{volume}{22}, \bibinfo{number}{2} (\bibinfo{year}{2017}),
  \bibinfo{pages}{579--630}.
\newblock
\showISSN{1573-7616}
\urldef\tempurl%
\url{https://doi.org/10.1007/s10664-016-9437-5}
\showDOI{\tempurl}


\bibitem[\protect\citeauthoryear{Klas, Trendowicz, Ishigai, and Nakao}{Klas
  et~al\mbox{.}}{2011}]%
        {klas_handling_2011}
\bibfield{author}{\bibinfo{person}{Michael Klas}, \bibinfo{person}{Adam
  Trendowicz}, \bibinfo{person}{Yasushi Ishigai}, {and} \bibinfo{person}{Haruka
  Nakao}.} \bibinfo{year}{2011}\natexlab{}.
\newblock \showarticletitle{Handling {Estimation} {Uncertainty} with
  {Bootstrapping}: {Empirical} {Evaluation} in the {Context} of {Hybrid}
  {Prediction} {Methods}}. In \bibinfo{booktitle}{\emph{Proc. of the {Int'l.}
  {Symposium} on {Empirical} {Software} {Engineering} and {Measurement}}}.
  \bibinfo{pages}{245--254}.
\newblock
\urldef\tempurl%
\url{https://doi.org/10.1109/ESEM.2011.33}
\showDOI{\tempurl}


\bibitem[\protect\citeauthoryear{Koru, Zhang, El~Emam, and Liu}{Koru
  et~al\mbox{.}}{2009}]%
        {koru_investigation_2009}
\bibfield{author}{\bibinfo{person}{Gunes Koru}, \bibinfo{person}{Dongsong
  Zhang}, \bibinfo{person}{Khaled El~Emam}, {and} \bibinfo{person}{Hongfang
  Liu}.} \bibinfo{year}{2009}\natexlab{}.
\newblock \showarticletitle{An {Investigation} into the {Functional} {Form} of
  the {Size}-{Defect} {Relationship} for {Software} {Modules}}.
\newblock \bibinfo{journal}{\emph{IEEE Trans. on Software Engineering}}
  \bibinfo{volume}{35}, \bibinfo{number}{2} (\bibinfo{year}{2009}),
  \bibinfo{pages}{293--304}.
\newblock
\showISSN{1939-3520}
\urldef\tempurl%
\url{https://doi.org/10.1109/TSE.2008.90}
\showDOI{\tempurl}


\bibitem[\protect\citeauthoryear{Lacerda, Petrillo, Pimenta, and
  Gueheneuc}{Lacerda et~al\mbox{.}}{2020}]%
        {lacerda_code_2020}
\bibfield{author}{\bibinfo{person}{Guilherme Lacerda}, \bibinfo{person}{Fabio
  Petrillo}, \bibinfo{person}{Marcelo Pimenta}, {and}
  \bibinfo{person}{Yann~Gael Gueheneuc}.} \bibinfo{year}{2020}\natexlab{}.
\newblock \showarticletitle{Code {Smells} and {Refactoring}: {A} {Tertiary}
  {Systematic} {Review} of {Challenges} and {Observations}}.
\newblock \bibinfo{journal}{\emph{Journal of Systems and Software}}
  \bibinfo{volume}{167} (\bibinfo{year}{2020}), \bibinfo{pages}{110610}.
\newblock


\bibitem[\protect\citeauthoryear{Lenarduzzi, Besker, Taibi, Martini, and
  Arcelli~Fontana}{Lenarduzzi et~al\mbox{.}}{2021}]%
        {lenarduzzi_systematic_2021}
\bibfield{author}{\bibinfo{person}{Valentina Lenarduzzi},
  \bibinfo{person}{Terese Besker}, \bibinfo{person}{Davide Taibi},
  \bibinfo{person}{Antonio Martini}, {and} \bibinfo{person}{Francesca
  Arcelli~Fontana}.} \bibinfo{year}{2021}\natexlab{}.
\newblock \showarticletitle{A {Systematic} {Literature} {Review} on {Technical}
  {Debt} {Prioritization}: {Strategies}, {Processes}, {Factors}, and {Tools}}.
\newblock \bibinfo{journal}{\emph{Journal of Systems and Software}}
  \bibinfo{volume}{171} (\bibinfo{year}{2021}), \bibinfo{pages}{110827}.
\newblock
\showISSN{0164-1212}
\urldef\tempurl%
\url{https://doi.org/10.1016/j.jss.2020.110827}
\showDOI{\tempurl}


\bibitem[\protect\citeauthoryear{Leven, Broman, Besker, and Torkar}{Leven
  et~al\mbox{.}}{2023}]%
        {leven_testing_2023}
\bibfield{author}{\bibinfo{person}{William Leven}, \bibinfo{person}{Hampus
  Broman}, \bibinfo{person}{Terese Besker}, {and} \bibinfo{person}{Richard
  Torkar}.} \bibinfo{year}{2023}\natexlab{}.
\newblock \bibinfo{title}{Testing the {Broken} {Windows} {Theory} in the
  context of {Technical} {Debt}}.
\newblock
\newblock
\urldef\tempurl%
\url{https://doi.org/10.48550/arXiv.2209.01549}
\showDOI{\tempurl}


\bibitem[\protect\citeauthoryear{Li and Shatnawi}{Li and Shatnawi}{2007}]%
        {li_empirical_2007}
\bibfield{author}{\bibinfo{person}{Wei Li} {and} \bibinfo{person}{Raed
  Shatnawi}.} \bibinfo{year}{2007}\natexlab{}.
\newblock \showarticletitle{An {Empirical} {Study} of the {Bad} {Smells} and
  {Class} {Error} {Probability} in the {Post}-{Release} {Object}-{Oriented}
  {System} {Evolution}}.
\newblock \bibinfo{journal}{\emph{Journal of Systems and Software}}
  \bibinfo{volume}{80}, \bibinfo{number}{7} (\bibinfo{year}{2007}),
  \bibinfo{pages}{1120--1128}.
\newblock
\showISSN{0164-1212}
\urldef\tempurl%
\url{https://doi.org/10.1016/j.jss.2006.10.018}
\showDOI{\tempurl}


\bibitem[\protect\citeauthoryear{Nunez-Varela, Perez-Gonzalez, Martinez-Perez,
  and Soubervielle-Montalvo}{Nunez-Varela et~al\mbox{.}}{2017}]%
        {nunez-varela_source_2017}
\bibfield{author}{\bibinfo{person}{Alberto~S. Nunez-Varela},
  \bibinfo{person}{Hector~G. Perez-Gonzalez}, \bibinfo{person}{Francisco~E.
  Martinez-Perez}, {and} \bibinfo{person}{Carlos Soubervielle-Montalvo}.}
  \bibinfo{year}{2017}\natexlab{}.
\newblock \showarticletitle{Source {Code} {Metrics}: {A} {Systematic} {Mapping}
  {Study}}.
\newblock \bibinfo{journal}{\emph{Journal of Systems and Software}}
  \bibinfo{volume}{128} (\bibinfo{year}{2017}), \bibinfo{pages}{164--197}.
\newblock
\showISSN{0164-1212}


\bibitem[\protect\citeauthoryear{Olbrich, Cruzes, Basili, and Zazworka}{Olbrich
  et~al\mbox{.}}{2009}]%
        {olbrich_evolution_2009}
\bibfield{author}{\bibinfo{person}{Steffen Olbrich},
  \bibinfo{person}{Daniela~S. Cruzes}, \bibinfo{person}{Victor Basili}, {and}
  \bibinfo{person}{Nico Zazworka}.} \bibinfo{year}{2009}\natexlab{}.
\newblock \showarticletitle{The {Evolution} and {Impact} of {Code} {Smells}:
  {A} {Case} {Study} of {Two} {Open} {Source} {Systems}}. In
  \bibinfo{booktitle}{\emph{Proc. of the 3rd {Int'l.} {Symposium} on
  {Empirical} {Software} {Engineering} and {Measurement}}}.
  \bibinfo{pages}{390--400}.
\newblock
\urldef\tempurl%
\url{https://doi.org/10.1109/ESEM.2009.5314231}
\showDOI{\tempurl}


\bibitem[\protect\citeauthoryear{Pearl}{Pearl}{2018}]%
        {pearl_book_2018}
\bibfield{author}{\bibinfo{person}{Judea Pearl}.}
  \bibinfo{year}{2018}\natexlab{}.
\newblock \bibinfo{booktitle}{\emph{The {Book} of {Why}: {The} {New} {Science}
  of {Cause} and {Effect}} (\bibinfo{edition}{1} ed.)}.
\newblock \bibinfo{publisher}{Ingram Publisher Services}, \bibinfo{address}{La
  Vergne, TN, USA}.
\newblock
\showISBNx{978-0-465-09760-9}


\bibitem[\protect\citeauthoryear{Radjenovic, Hericko, Torkar, and
  Zivkovic}{Radjenovic et~al\mbox{.}}{2013}]%
        {radjenovic_software_2013}
\bibfield{author}{\bibinfo{person}{Danijel Radjenovic}, \bibinfo{person}{Marjan
  Hericko}, \bibinfo{person}{Richard Torkar}, {and} \bibinfo{person}{Ales
  Zivkovic}.} \bibinfo{year}{2013}\natexlab{}.
\newblock \showarticletitle{Software {Fault} {Prediction} {Metrics}: {A}
  {Systematic} {Literature} {Review}}.
\newblock \bibinfo{journal}{\emph{Information and Software Technology}}
  \bibinfo{volume}{55}, \bibinfo{number}{8} (\bibinfo{year}{2013}),
  \bibinfo{pages}{1397--1418}.
\newblock
\showISSN{0950-5849}
\urldef\tempurl%
\url{https://doi.org/10.1016/j.infsof.2013.02.009}
\showDOI{\tempurl}


\bibitem[\protect\citeauthoryear{Riaz, Mendes, and Tempero}{Riaz
  et~al\mbox{.}}{2009}]%
        {riaz_systematic_2009}
\bibfield{author}{\bibinfo{person}{Mehwish Riaz}, \bibinfo{person}{Emilia
  Mendes}, {and} \bibinfo{person}{Ewan Tempero}.}
  \bibinfo{year}{2009}\natexlab{}.
\newblock \showarticletitle{A {Systematic} {Review} of {Software}
  {Maintainability} {Prediction} and {Metrics}}. In
  \bibinfo{booktitle}{\emph{Proc. of the 3rd {Int}'l. {Symposium} on
  {Empirical} {Software} {Engineering} and {Measurement}}}.
  \bibinfo{pages}{367--377}.
\newblock


\bibitem[\protect\citeauthoryear{Rios, Mendonca~Neto, and Spinola}{Rios
  et~al\mbox{.}}{2018}]%
        {rios_tertiary_2018}
\bibfield{author}{\bibinfo{person}{Nicolli Rios}, \bibinfo{person}{Manoel
  Gomes~de Mendonca~Neto}, {and} \bibinfo{person}{Rodrigo~Oliveira Spinola}.}
  \bibinfo{year}{2018}\natexlab{}.
\newblock \showarticletitle{A {Tertiary} {Study} on {Technical} {Debt}:
  {Types}, {Management} {Strategies}, {Research} {Trends}, and {Base}
  {Information} for {Practitioners}}.
\newblock \bibinfo{journal}{\emph{Information and Software Technology}}
  \bibinfo{volume}{102} (\bibinfo{year}{2018}), \bibinfo{pages}{117--145}.
\newblock
\showISSN{0950-5849}


\bibitem[\protect\citeauthoryear{Santos, Rocha-Junior, Prates, Nascimento,
  Freitas, and Mendonca}{Santos et~al\mbox{.}}{2018}]%
        {santos_systematic_2018}
\bibfield{author}{\bibinfo{person}{Jose Amancio~M. Santos},
  \bibinfo{person}{Joao~B. Rocha-Junior}, \bibinfo{person}{Luciana Carla~Lins
  Prates}, \bibinfo{person}{Rogeres Santos~do Nascimento},
  \bibinfo{person}{Mydia~Falcao Freitas}, {and} \bibinfo{person}{Manoel
  Gomes~de Mendonca}.} \bibinfo{year}{2018}\natexlab{}.
\newblock \showarticletitle{A {Systematic} {Review} on the {Code} {Smell}
  {Effect}}.
\newblock \bibinfo{journal}{\emph{Journal of Systems and Software}}
  \bibinfo{volume}{144} (\bibinfo{year}{2018}), \bibinfo{pages}{450--477}.
\newblock
\showISSN{0164-1212}
\urldef\tempurl%
\url{https://doi.org/10.1016/j.jss.2018.07.035}
\showDOI{\tempurl}


\bibitem[\protect\citeauthoryear{Sjoberg and Bergersen}{Sjoberg and
  Bergersen}{2023}]%
        {sjoberg_improving_2023}
\bibfield{author}{\bibinfo{person}{Dag Sjoberg} {and}
  \bibinfo{person}{Gunnar~Rye Bergersen}.} \bibinfo{year}{2023}\natexlab{}.
\newblock \showarticletitle{Improving the {Reporting} of {Threats} to
  {Construct} {Validity}}. In \bibinfo{booktitle}{\emph{Proc. of the 27th
  {Int'l.} {Conf.} on {Evaluation} and {Assessment} in {Software}
  {Engineering}}}. \bibinfo{pages}{205--209}.
\newblock
\showISBNx{9798400700446}
\urldef\tempurl%
\url{https://doi.org/10.1145/3593434.3593449}
\showDOI{\tempurl}


\bibitem[\protect\citeauthoryear{Sjoberg, Yamashita, Anda, Mockus, and
  Dyba}{Sjoberg et~al\mbox{.}}{2013}]%
        {sjoberg_quantifying_2013}
\bibfield{author}{\bibinfo{person}{Dag~I.K. Sjoberg}, \bibinfo{person}{Aiko
  Yamashita}, \bibinfo{person}{Bente~C.D. Anda}, \bibinfo{person}{Audris
  Mockus}, {and} \bibinfo{person}{Tore Dyba}.} \bibinfo{year}{2013}\natexlab{}.
\newblock \showarticletitle{Quantifying the {Effect} of {Code} {Smells} on
  {Maintenance} {Effort}}.
\newblock \bibinfo{journal}{\emph{IEEE Trans. on Software Engineering}}
  \bibinfo{volume}{39}, \bibinfo{number}{8} (\bibinfo{year}{2013}),
  \bibinfo{pages}{1144--1156}.
\newblock
\showISSN{1939-3520}
\urldef\tempurl%
\url{https://doi.org/10.1109/TSE.2012.89}
\showDOI{\tempurl}


\bibitem[\protect\citeauthoryear{Soh, Yamashita, Khomh, and Gueheneuc}{Soh
  et~al\mbox{.}}{2016}]%
        {soh_code_2016}
\bibfield{author}{\bibinfo{person}{Zephyrin Soh}, \bibinfo{person}{Aiko
  Yamashita}, \bibinfo{person}{Foutse Khomh}, {and} \bibinfo{person}{Yann-Gael
  Gueheneuc}.} \bibinfo{year}{2016}\natexlab{}.
\newblock \showarticletitle{Do {Code} {Smells} {Impact} the {Effort} of
  {Different} {Maintenance} {Programming} {Activities}?}. In
  \bibinfo{booktitle}{\emph{Proc. of the 23rd {Int'l.} {Conf.} on {Software}
  {Analysis}, {Evolution}, and {Reengineering}}}, Vol.~\bibinfo{volume}{1}.
  \bibinfo{pages}{393--402}.
\newblock
\urldef\tempurl%
\url{https://doi.org/10.1109/SANER.2016.103}
\showDOI{\tempurl}


\bibitem[\protect\citeauthoryear{Tawosi, Moussa, and Sarro}{Tawosi
  et~al\mbox{.}}{2022}]%
        {tawosi_relationship_2022}
\bibfield{author}{\bibinfo{person}{Vali Tawosi}, \bibinfo{person}{Rebecca
  Moussa}, {and} \bibinfo{person}{Federica Sarro}.}
  \bibinfo{year}{2022}\natexlab{}.
\newblock \showarticletitle{On the {Relationship} {Between} {Story} {Points}
  and {Development} {Effort} in {Agile} {Open}-{Source} {Software}}. In
  \bibinfo{booktitle}{\emph{Proc. of the 16th {Int'l.} {Symposium} on
  {Empirical} {Software} {Engineering} and {Measurement}}}.
  \bibinfo{pages}{183--194}.
\newblock
\showISBNx{978-1-4503-9427-7}


\bibitem[\protect\citeauthoryear{Tornhill}{Tornhill}{2018}]%
        {tornhill_software_2018}
\bibfield{author}{\bibinfo{person}{Adam Tornhill}.}
  \bibinfo{year}{2018}\natexlab{}.
\newblock \bibinfo{booktitle}{\emph{Software {Design} {X}-{Rays}: {Fix}
  {Technical} {Debt} with {Behavioral} {Code} {Analysis}}}.
\newblock \bibinfo{publisher}{Pragmatic Bookshelf}.
\newblock


\bibitem[\protect\citeauthoryear{Tornhill and Borg}{Tornhill and Borg}{2022}]%
        {tornhill_code_2022}
\bibfield{author}{\bibinfo{person}{Adam Tornhill} {and} \bibinfo{person}{Markus
  Borg}.} \bibinfo{year}{2022}\natexlab{}.
\newblock \showarticletitle{Code {Red}: {The} {Business} {Impact} of {Code}
  {Quality} - {A} {Quantitative} {Study} of 39 {Proprietary} {Production}
  {Codebases}}. In \bibinfo{booktitle}{\emph{Proc. of the 5th {Int'l.} {Conf.}
  on {Technical} {Debt}}}. \bibinfo{pages}{11--20}.
\newblock


\bibitem[\protect\citeauthoryear{Tufekci}{Tufekci}{2022}]%
        {tufekci_shameful_2022}
\bibfield{author}{\bibinfo{person}{Zeynep Tufekci}.}
  \bibinfo{year}{2022}\natexlab{}.
\newblock \showarticletitle{The {Shameful} {Open} {Secret} {Behind}
  {Southwest}’s {Failure}}.
\newblock \bibinfo{journal}{\emph{The New York Times}} (\bibinfo{date}{Dec.}
  \bibinfo{year}{2022}).
\newblock
\showISSN{0362-4331}
\urldef\tempurl%
\url{https://www.nytimes.com/2022/12/31/opinion/southwest-airlines-computers.html}
\showURL{%
\tempurl}


\bibitem[\protect\citeauthoryear{Wilson and Kelling}{Wilson and
  Kelling}{1982}]%
        {wilson_broken_1982}
\bibfield{author}{\bibinfo{person}{James Wilson} {and} \bibinfo{person}{George
  Kelling}.} \bibinfo{year}{1982}\natexlab{}.
\newblock \showarticletitle{Broken {Windows}: {The} {Police} and {Neighborhood}
  {Safety}}.
\newblock \bibinfo{journal}{\emph{Atlantic}} (\bibinfo{year}{1982}),
  \bibinfo{pages}{29--38}.
\newblock


\bibitem[\protect\citeauthoryear{Yamashita and Moonen}{Yamashita and
  Moonen}{2012}]%
        {yamashita_code_2012}
\bibfield{author}{\bibinfo{person}{Aiko Yamashita} {and} \bibinfo{person}{Leon
  Moonen}.} \bibinfo{year}{2012}\natexlab{}.
\newblock \showarticletitle{Do {Code} {Smells} {Reflect} {Important}
  {Maintainability} {Aspects}?}. In \bibinfo{booktitle}{\emph{Proc. of the 28th
  {IEEE} {Int'l.} {Conf.} on {Software} {Maintenance}}}.
  \bibinfo{pages}{306--315}.
\newblock
\urldef\tempurl%
\url{https://doi.org/10.1109/ICSM.2012.6405287}
\showDOI{\tempurl}


\bibitem[\protect\citeauthoryear{Zazworka, Shaw, Shull, and Seaman}{Zazworka
  et~al\mbox{.}}{2011}]%
        {zazworka_investigating_2011}
\bibfield{author}{\bibinfo{person}{Nico Zazworka}, \bibinfo{person}{Michele~A.
  Shaw}, \bibinfo{person}{Forrest Shull}, {and} \bibinfo{person}{Carolyn
  Seaman}.} \bibinfo{year}{2011}\natexlab{}.
\newblock \showarticletitle{Investigating the {Impact} of {Design} {Debt} on
  {Software} {Quality}}. In \bibinfo{booktitle}{\emph{Proc. of the 2nd
  {Workshop} on {Managing} {Technical} {Debt}}}. \bibinfo{pages}{17--23}.
\newblock
\showISBNx{978-1-4503-0586-0}
\urldef\tempurl%
\url{https://doi.org/10.1145/1985362.1985366}
\showDOI{\tempurl}


\bibitem[\protect\citeauthoryear{Zhang, Zhou, and Zhu}{Zhang
  et~al\mbox{.}}{2017}]%
        {zhang_empirical_2017}
\bibfield{author}{\bibinfo{person}{Xiaofang Zhang}, \bibinfo{person}{Yida
  Zhou}, {and} \bibinfo{person}{Can Zhu}.} \bibinfo{year}{2017}\natexlab{}.
\newblock \showarticletitle{An {Empirical} {Study} of the {Impact} of {Bad}
  {Designs} on {Defect} {Proneness}}. In \bibinfo{booktitle}{\emph{Proc. of the
  {Int'l.} {Conf.} on {Software} {Analysis}, {Testing} and {Evolution}}}.
  \bibinfo{pages}{1--9}.
\newblock
\urldef\tempurl%
\url{https://doi.org/10.1109/SATE.2017.9}
\showDOI{\tempurl}


\end{thebibliography}

\end{document}